\documentclass[11pt,twoside,a4paper]{article}
\pdfoutput=1

\usepackage{amsmath,amsfonts,amssymb}
\usepackage{marginnote}

\numberwithin{equation}{section}

\usepackage{cancel}
\usepackage{cite}

\usepackage[pdftex]{color,graphicx}
\usepackage[english]{babel}
\usepackage[latin9]{inputenc}

\usepackage{color}
\usepackage{multirow}
\usepackage{enumerate}
\usepackage{mdwlist}
\usepackage{verbatim}

\def\2b2[#1,#2][#3,#4]{\left( \begin{array}{cc} #1 & #2 \\ #3 & #4 \end{array} \right)}
\def\arrb3[#1,#2,#3][#4,#5,#6][#7,#8,#9]{\left( \begin{array}{ccc} #1 & #2 & #3 \\ #4 & #5 & #6\\#7&#8&#9\end{array} \right)}
\def\6diag[#1,#2,#3]{\left( \begin{array}{cccccc} #1 & 0 & 0 & 0 & 0 & 0 \\ 0 & #1 & 0 & 0 & 0 & 0 \\ 0 & 0& #2 & 0 & 0 &0 \\ 0 & 0 & 0 & #2 & 0 & 0 \\ 0 & 0 & 0 & 0 & #3 & 0 \\ 0 & 0 & 0 & 0 & 0 & #3  \end{array} \right)}
\def\3diag[#1,#2,#3]{\left(\begin{array}{ccc} #1 & 0 & 0 \\ 0 & #2 &0\\ 0&0&#3 \end{array}\right)}

\def\arr2[#1,#2][#3,#4]{\left( \begin{array}{cc} #1 & #2 \\ #3 & #4 \end{array} \right)}
\def\vec2[#1,#2]{\left( \begin{array}{c} #1 \\ #2 \end{array} \right)}
\def\vect3[#1,#2,#3]{\left( \begin{array}{c} #1 \\ #2 \\ #3\end{array} \right)}

\newcommand{\U}[1]{\ensuremath{\mathrm{U}(#1)}}
\newcommand{\unit}[1]{\; \mathrm{#1}}

\newcommand{\GeV}{\unit{GeV}}

\newcommand{\cm}{\unit{cm}}

\newcommand{\sqcm}{\cm^{2}}

\newcommand{\s}{\unit{s}}

\def\bra{\langle}
\def\ket{\rangle}

\def\beq{\begin{equation}}
\def\eeq{\end{equation}}
\def\bal{\begin{align}}
\def\eal{\end{align}}
\def\nn{\nonumber}

\newcommand{\bea}{\begin{eqnarray}}
\newcommand{\eea}{\end{eqnarray}}

\newcommand{\C}[1]{\mathcal{#1}}

\def\ov{\overline}

\date{}

\begin{document}

\title{\vspace{-3cm}
\hfill{\small{DESY 11-159}}\\[2cm]
{\bf \LARGE Dark Matter and Dark Forces from a supersymmetric
hidden sector}}

\author{S.~Andreas\footnote{sarah.andreas@desy.de},
M.~D.~Goodsell\footnote{mark.goodsell@desy.de, mark.goodsell@cern.ch},
A.~Ringwald\footnote{andreas.ringwald@desy.de} }

\maketitle \vspace{-1cm}

\begin{center}
\emph{Deutsches  Elektronen-Synchrotron, DESY, Notkestra\ss e 85, 22607  Hamburg, Germany}
\end{center}

\thispagestyle{empty}

\begin{abstract}
We show that supersymmetric ``Dark Force'' models with gravity
mediation are viable. To this end, we analyse a simple string-inspired
supersymmetric hidden sector model that interacts with the
visible sector via kinetic mixing of a light Abelian gauge
boson with the hypercharge. We include all induced interactions
with the visible sector such as neutralino mass mixing and the
Higgs portal term. We perform a detailed parameter space scan
comparing the produced dark matter relic abundance and direct
detection cross sections to current experiments.
\end{abstract}

\vspace{2cm}

\newpage

\section{Introduction}
\label{SEC:INTRODUCTION}

There has been much interest recently in the possibility that
there exists a hidden sector containing a dark matter particle
coupled to a hidden \U1 gauge boson (a ``Dark Force'') having a
mass of the order of a GeV that kinetically mixes with the
photon~\cite{Feldman:2006wd,Pospelov:2007mp,ArkaniHamed:2008qn,Pospelov:2008jd,ArkaniHamed:2008qp}.
Such a scenario could explain many astrophysical puzzles, such
as the  positron excess observed by
PAMELA~\cite{Adriani:2008zr}, ATIC~\cite{:2008zzr}, and
Fermi~\cite{Abdo:2009zk}, or the direct detection and annual
modulation signals of DAMA~\cite{Bernabei:2008yi},
CoGeNT~\cite{Aalseth:2010vx,Aalseth:2011wp} and
CRESST~\cite{Angloher:2011uu} (if one ignores the
disputed~\cite{Collar:2011wq,Collar:2011kf} contradiction due
to XENON100~\cite{Aprile:2011hi} and CDMS~\cite{Ahmed:2010wy}).
Following from the work
of~\cite{Chun:2008by,Cheung:2009qd,Katz:2009qq}, elegantly
simple supersymmetric models in the latter category were
constructed in~\cite{Morrissey:2009ur} and further examined
in~\cite{Cohen:2010kn} (see also~\cite{Kang:2010mh}). However,
these works emphasized that, in order to obtain such a light
hidden sector, supersymmetry breaking effects in the visible
sector would necessarily be dominated by gauge mediation, in
order that the masses should be acceptably small. Thus it is
natural to ask whether confirmation of these observations would
be in contradiction with gravity mediation; in other words,
whether it is also possible to have a gravity-mediated spectrum
of particles that can yield similar phenomenology. This is also
linked to the interesting question as to whether these models
can be embedded into string theory: such hidden sectors appear
very naturally there -- see,
e.g.,~\cite{Abel:2008ai,Goodsell:2009pi,Goodsell:2009xc,Goodsell:2010ie,Heckman:2010fh,Bullimore:2010aj,Cicoli:2011yh,Williams:2011qb,Heckman:2011sw}
-- but the problem of finding gauge mediation dominance over
gravity mediation is notoriously difficult to achieve in
globally consistent models.

Beyond the dark matter motivation, it is also useful to ask what
hidden sector models of this form coming from string theory are
allowed or excluded by current observations. This is because,
even if the hidden sector does not comprise (all) the dark
matter, there is a wealth of experiments capable of probing
Dark Forces over a very wide range of hidden gauge boson mass
and kinetic mixing values. Kinetic mixing was considered in the
context of the heterotic string
in~\cite{Dienes:1996zr,Lukas:1999nh,Blumenhagen:2005ga,Goodsell:2010ie,Goodsell:2011wn}.
It has been examined in type II strings
in~\cite{Abel:2003ue,Lust:2003ky,Abel:2006qt,Abel:2008ai,Benakli:2009mk,Goodsell:2009xc,Gmeiner:2009fb,Goodsell:2009pi};
in~\cite{Goodsell:2009xc,Cicoli:2011yh}, both masses and mixings
were considered, and it was argued that the Dark Forces
scenario could be accommodated provided that there is
additional sequestering. In this work, we shall consider hidden
sector models with the particle content and similar couplings
to those in~\cite{Morrissey:2009ur}, but argue that when we
have gravity mediation domination, these can still give
interesting phenomenological predictions under certain mild
assumptions, without requiring additional sequestering relative
to the visible sector. Although we will discuss the possible
explanation of the signals found by DAMA and CoGeNT, these will
therefore not be our primary motivation: rather, we wish to
explore how simple supersymmetric hidden dark sectors with a
hidden \U1 can be constrained by observations.

The paper is organised as follows. In section~\ref{SEC:SUSYDS},
we describe the model of a supersymmetric dark sector that we
shall be examining. This is followed by a summary of
constraints upon hidden \U1s with hidden matter charged under
them in section~\ref{SEC:CONSTRAINTS}. There we also include
the reach of future fixed target experiments and illustrate
these with an investigation of a simple toy model.
Section~\ref{SEC:RESULTS} then contains the meat of the paper:
the results of the parameter search over our supersymmetric
dark sector model. We include additional technical details in
the appendix: the hidden sector renormalisation
group equations (RGEs) in
appendix~\ref{APP:RGEs}; the spectrum of the model in
appendix~\ref{APP:LowE} (including the mass mixing matrix with
the visible neutralino in~\ref{APP:FERMIONS}); a review of
kinetic and mass mixing of a massive hidden gauge boson with
the hypercharge and $Z$ in appendix~\ref{APP:KM}; and a
description of the Goldstone boson mixing in
appendix~\ref{APP:GOLDSTONE}. In addition, in
appendix~\ref{APP:HIGGS}, we discuss the supersymmetry-induced
Higgs portal term and the mixing of the hidden and minimal supersymmetric standard model (MSSM) Higgs
fields; we believe that although the existence of the term has
been known in the literature (see, e.g.,~\cite{Foot:1991bp} in non-SUSY models and \cite{Schabinger:2005ei} in the SUSY context) the effect of the mixing terms for direct detection have not been
given elsewhere. Included is a calculation of the
induced coupling of the hidden dark matter Majorana fermion to
nucleons.

\section{Supersymmetric dark sectors}
\label{SEC:SUSYDS}

\subsection{Supersymmetric kinetic mixing}

We shall consider models that interact with the visible sector
primarily through kinetic mixing of a hidden \U1 gauge field
with the hypercharge. Hence, we have a holomorphic kinetic
mixing $\chi_h$ between hypercharge $B_\alpha$ with coupling
$g_Y$ (and gaugino the Bino, $b$) and hidden gauge superfield
$X_\alpha$ with coupling $g_h$ (and gaugino written as
$\lambda$) appearing in the Lagrangian density
\begin{equation}
\C{L} \supset \int d^2 \theta \bigg(\frac{1}{4g_Y^2} B^\alpha B_\alpha + \frac{1}{4g_h^2} X^\alpha
X_\alpha - \frac{\chi_h}{2} B^\alpha X_\alpha \bigg)\,.
\label{EQ:CANONICAL}\end{equation}
The physical kinetic mixing in the canonical basis
\cite{Benakli:2009mk,Goodsell:2009xc} is then given by
\begin{equation}
\chi = g_Y g_h \mathrm{Re} (\chi_h) .
\end{equation}
We shall assume no matter charged under both hidden and visible
gauge groups, so this relationship is valid at all energy
scales.  Since we are considering string-inspired models with a
``hidden'' \U1, that is, without matter charged under both the
visible and hidden gauge groups, we shall take the value of the
holomorphic kinetic mixing parameter to be of the order of a
loop factor \cite{Goodsell:2009xc}:
\begin{align}
\chi_h \equiv& \frac{\kappa}{16\pi^2}.
\end{align}
Here, $\kappa$ is a number that must, in principle, be derived from the high-energy model; in a field theory model, it is generated by integrating out some heavy linking fields (charged under visible and hidden sectors) at one loop, whereas in string models, it can be understood as arising from Kaluza-Klein modes of closed strings. In all cases, it depends only logarithmically upon mass splittings of the spectrum, and we shall therefore either take it to be equal to one or to vary by at most an order of magnitude from unity~\cite{Goodsell:2009xc,Bullimore:2010aj,Cicoli:2011yh}\footnote{Our results only depend on the absolute value of the mixing parameter. Effects that are sensitive to the different signs have been studied in~\cite{Feldman:2007wj}.}. We thus have
\begin{equation}
\chi =  g_Y g_h \frac{\kappa}{16\pi^2};
\label{ChiRelation}
\end{equation}
the most commonly taken value for $\chi$ is thus of the order of $10^{-3}$, but smaller values correspond to decreasing the hidden gauge coupling which may be extremely small in the case of hyperweak groups \cite{Burgess:2008ri,Goodsell:2009xc,Cicoli:2011yh}. Henceforth, we shall always use the physical mixing $\chi$.

As befits a well-studied subject, there are a variety of notations. In addition to using $\chi$, we shall also adopt the notation used in \cite{Hook:2010tw}:\footnote{Note, however, that this differs from the expressions in~\cite{Morrissey:2009ur}, which defines $\chi = -\tilde{\epsilon} \equiv -  t_{\tilde{\epsilon}}, s_{\tilde{\epsilon}} \equiv  -\frac{\chi}{\sqrt{1-\chi^2}}, c_{\tilde{\epsilon}} \equiv  \frac{1}{\sqrt{1-\chi^2}}$, although there they write $\epsilon$ instead of $\tilde{\epsilon}$ (we added the tilde to avoid confusion with the above). On the other hand, \cite{Feldman:2007wj,Dudas:2009uq,Mambrini:2009ad,Mambrini:2010yp,Mambrini:2010dq,Mambrini:2011dw} define $\delta \equiv - \chi$.}
\begin{align}
\chi \equiv& - \sin \epsilon \equiv  - s_\epsilon\nn\\
\cos \epsilon \equiv c_\epsilon \equiv  \sqrt{1-\chi^2},& \qquad \tan \epsilon \equiv t_\epsilon = - \frac{\chi}{\sqrt{1-\chi^2}}.
\end{align}
However, a crucial novelty in this work is the application of relation (\ref{ChiRelation}) to parameter scans rather than allowing for independent $\chi$ and $g_h$, which we shall see in section~\ref{SEC:TOY} will lead to qualitatively different results for the cross sections.

\subsection{Hidden matter fields}

The model that we shall consider is the simplest possible
without adding dimensionful supersymmetric quantities. There
are three chiral superfields $S, H_+, H_-$ with $H_+$ and $H_-$
charged under the hidden \U1 with charges $\pm 1$. These appear
in a superpotential with dimensionless coupling~$\lambda_S$
\beq W \supset \lambda_S S H_+ H_-. \eeq This is inspired by
$D$-brane models where the singlet is essentially the adjoint of
the gauge group: the superpotential above arises due to the
$N=2$-like structure, and there is no renormalisable singlet
potential due to this; alternatively, there may be $N=2$
supersymmetry of the couplings at some scale, although we shall
not enforce this. Such hidden sectors from string theory were
considered in, e.g.,~\cite{Heckman:2010fh,Heckman:2011sw}, and
the above model was studied with gauge mediation
in~\cite{Morrissey:2009ur} where it was termed a ``hidden
sector NMSSM,'' although we have set the cubic singlet term in
the superpotential to zero. There then exists a global \U1
symmetry under which $S$ and $H_-$ are charged; string theory
will not respect this, and we consider that it shall either be
broken at higher order in the superpotential or through
non-perturbative effects -- but we shall assure that it will
play no role in the following.

Once we include soft supersymmetry-breaking terms, we have the
approximate potential for the hidden sector,
\begin{align}
V =& |\lambda_S|^2 ( |S H_+|^2 + |S H_-|^2 + |H_+ H_-|^2) \nn\\
&+\frac{g_h^2}{2} (|H_+|^2 - |H_-|^2 - \xi)^2  \nn\\
&+m_+^2 |H_+|^2 + m_-^2 |H_-|^2 + m_S^2 |S|^2 \nn\\
&+ (\lambda_S A_S SH_+ H_- + \frac{1}{2} M_\lambda \lambda \lambda + c.c.),
\end{align}
where $\xi = - \frac{\chi}{g_h} \xi_Y = \chi (g_Y/g_h) g_Y
\frac{v^2}{4} \cos 2\beta$. The approximation lies in the
$D$-term potential; the full form is found in
appendix~\ref{APP:HIGGS}.

A crucial difference for the phenomenology of the model once we
consider gravity mediation is, however, that the gravitino is
not the lightest supersymmetric particle (LSP), and therefore the dark matter can consist of
stable hidden sector particles. We can thus perform a full
analysis of the model, including the visible sector and its
couplings, using
micrOMEGAs~\cite{Belanger:2001fz,Belanger:2006is,Belanger:2007zz,Belanger:2008sj,Belanger:2010pz}
to determine the relic abundance and direct detection cross
sections.

\subsection{Symmetry breaking through running}

Just as in the MSSM, the top Yukawa coupling can, through
running from the grand unified theory (GUT) scale, induce electroweak symmetry
breaking, so in the model we are considering, the Yukawa
coupling $\lambda_S$ can induce breaking of the hidden gauge
symmetry. By choosing the soft masses and couplings at the MSSM
GUT scale we can then find models at the low-energy scale with
hidden gauge symmetry breaking. \textit{A priori} the independent
supersymmetric parameters are $\chi, g_h, \lambda_S$ and the
soft masses $m_{H_\pm}, m_S, A_S$ and $M_\lambda$ (the hidden
gaugino mass) which we can choose at the high-energy scale and
run down.

Via (\ref{ChiRelation}), we are asserting a relation between
$\chi$ and $g_h$. Thus, if we take $\kappa =1$, we reduce the
number of free parameters in the model by one. However, as
described above, we shall in certain plots (figures \ref{Fig:ToyModPlot},\ref{RadScanNoSIMPLE} right,\ref{RadScanWithSIMPLE} right,\ref{RadScat},\ref{VisibleShowKappa} right,\ref{VisibleShowHalos},\ref{VisibleSDMajorana} right and \ref{VisibleSI}) allow an order of  %\marginnote{fig3,5r,6r,7,\\8r,9,10r,11} %
magnitude variation in $\kappa$; hence, although this does not
strictly reduce the number of parameters in the model, it does
rather constrain them with important consequences. Finally, we
shall make one further assumption about the parameters: we
shall take $m_{H_+} = m_{H_-}$ at the high-energy scale. This
is motivated by the fields $H_{\pm}$ being a non-chiral pair
(note that we are taking no explicit Fayet-Iliopoulos term for
the hidden $U(1)$ which would introduce a mass splitting).
Otherwise, we shall scan over the remaining parameters to find
interesting models.

\begin{figure}[htb!]
\begin{center}
\includegraphics[width=0.49\textwidth]{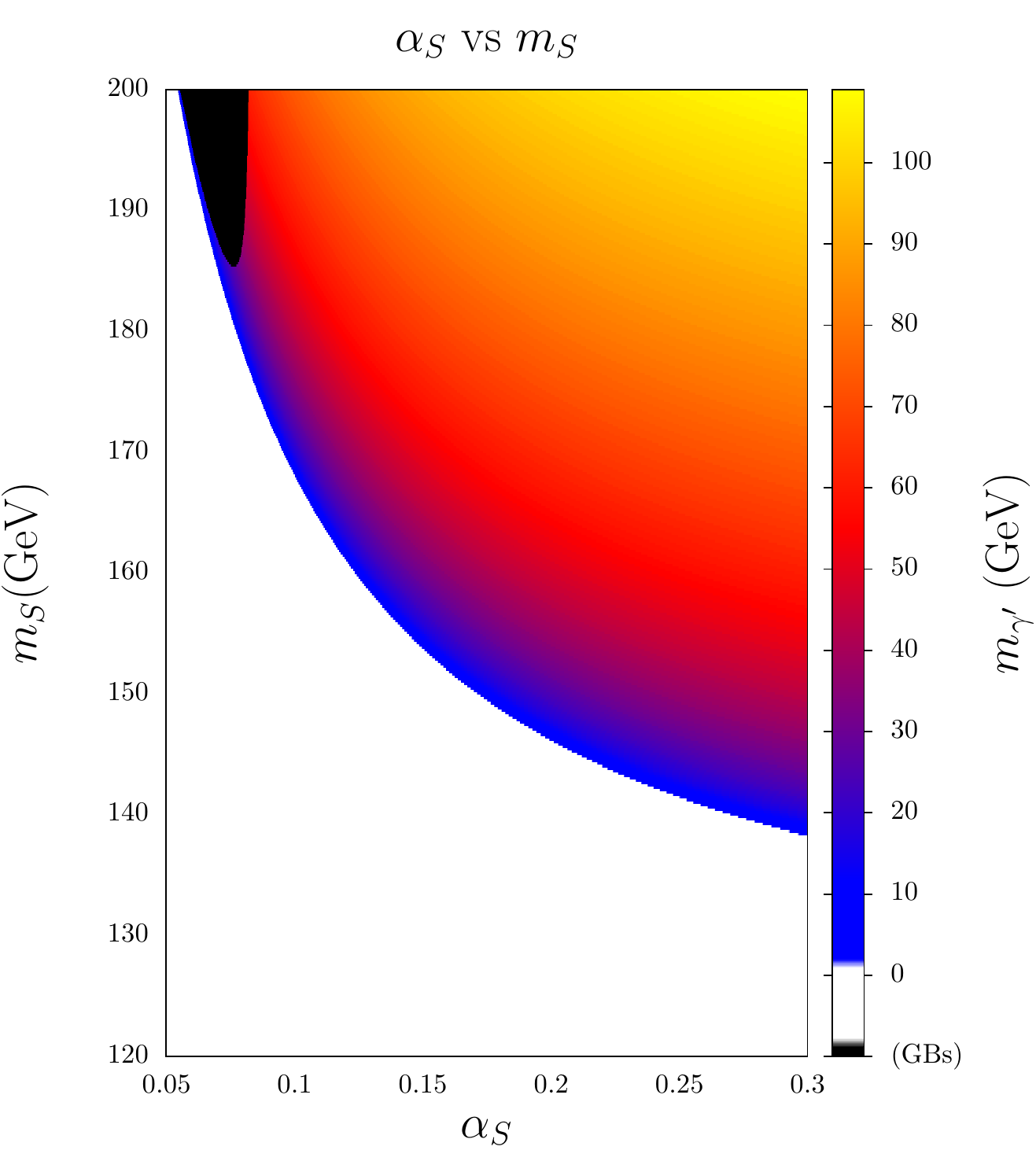}
\includegraphics[width=0.49\textwidth]{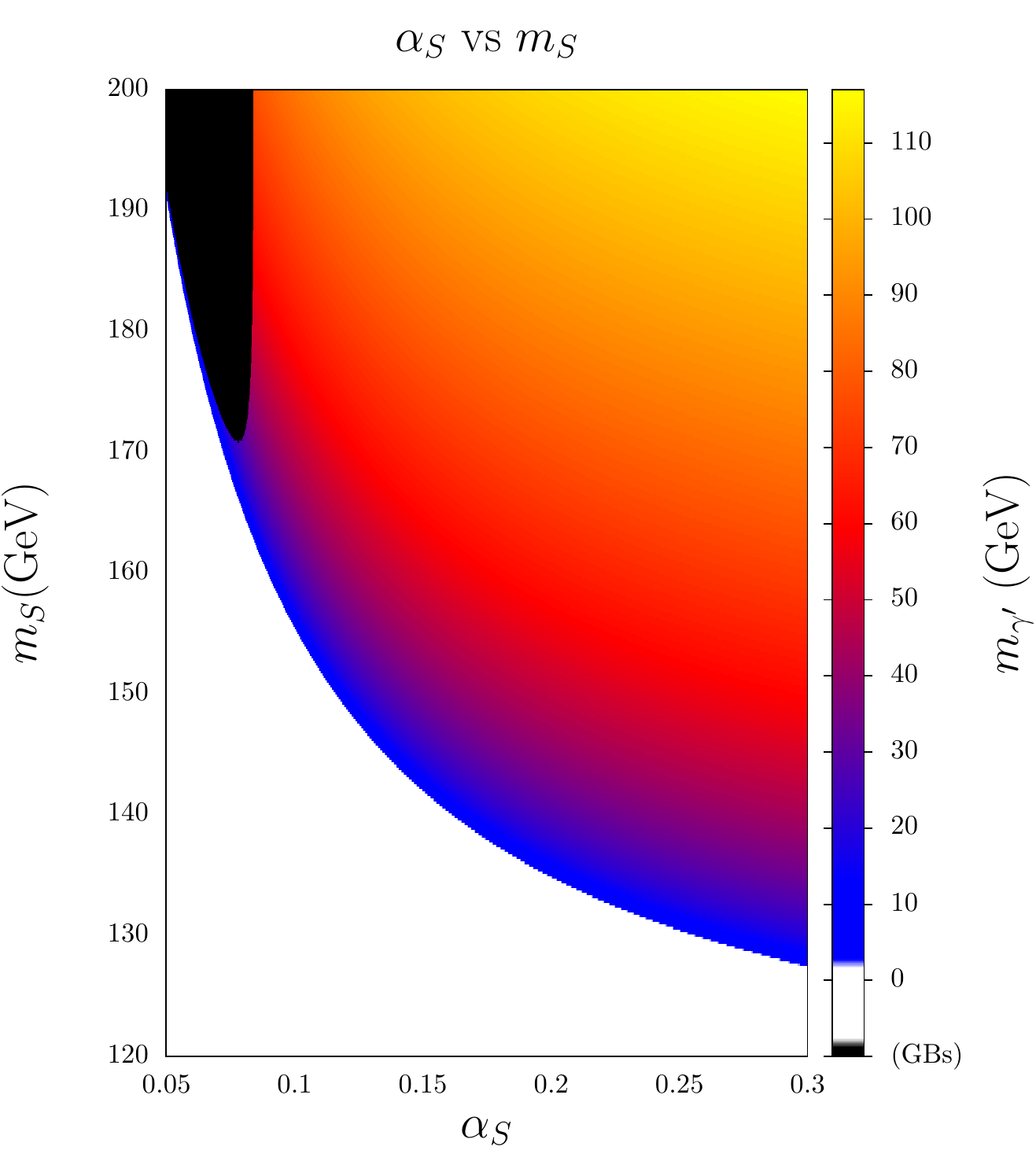}
%\marginnote{\hspace{-1cm}size increased}
\caption{Hidden photon mass $m_{\gamma'}$ induced by radiative hidden gauge
symmetry breaking, scanned over $m_S$ and
$\alpha_S \equiv \frac{\lambda_S^2}{4\pi}$. In both,
$m_H = A_S =100$ GeV, $\alpha_h= 0.0417$. \textit{Left:}
$M_\lambda = 71$ GeV, \textit{right:}  $M_\lambda = 50$ GeV.
All values given at $10^{16}$ GeV. The black region shows no stable symmetry breaking.}
\label{FIG:U1massesrun}
\end{center}
\end{figure}

The two-loop RGEs for the model are given in appendix
\ref{APP:RGEs}. By taking $m_S > m_{H_\pm}$ at the high-energy
scale, the RGEs naturally drive the soft masses for
$m_{H_\pm}^2$ to be negative at low energies, triggering hidden
symmetry breaking.\footnote{We ignore the effect on the running
of the kinetic mixing, since such terms always enter suppressed
by $\C{O}(\chi^2)$~\cite{Morrissey:2009ur} with an additional loop factor -- and are thus equivalent to \emph{three}-loop order.
Of course, it would be interesting to include all of these effects, where then the hidden
sector running would then be (extremely weakly) dependent
upon the visible sector parameters, and we leave this to future
work.} The visible sector coupling via the kinetic mixing then
determines which field ($H_+$ or $H_-$) condenses; without loss
of generality, we take $\chi$ to be negative, and thus $H_+$
condenses.  Defining $\Delta \equiv \sqrt{\lambda_S^2 \xi -
m_+^2 \lambda_S^2/g_h^2}$, we have the conditions for a stable
minimum with $\bra H_+ \ket = \Delta/\lambda_S$ and all other
expectation values zero:
\begin{align}
0 \le& \Delta^2 \nn\\
0 \le& m_-^2 + m_+^2 + m_S^2 + 2\Delta^2 \nn\\
0 \le& ( m_-^2 + m_+^2 + \Delta^2) ( m_S^2 + \Delta^2 ) - |A_S|^2 \Delta^2.
\end{align}
This is reviewed in appendix~\ref{APP:LowE}. The hidden gauge
boson mass is then given by
\begin{equation}
m_{\gamma'} =(\sqrt{2}g_h/\lambda_S) \Delta.
\end{equation}
We give two examples of the values obtained scanning over $m_S$
and $\alpha_S \equiv \frac{\lambda_S^2}{4\pi}$ in
figure~\ref{FIG:U1massesrun}.

\subsection{Symmetry breaking induced by the visible sector}

The mechanism for hidden gauge symmetry breaking promoted in
work such as \cite{Morrissey:2009ur} is via the effective
Fayet-Iliopoulos term induced in the hidden sector by the
kinetic mixing with the visible Higgs $D$-term. In such a case,
the mass-squareds $m_+^2, m_-^2$ may be positive provided they
are small enough that $\Delta^2 > 0$.

One motivation for this
work is that such a case is more difficult to justify in the case of
gravity mediation, but it is  not implausible, since it can be achieved, for example,
through  sequestering of the hidden sector. In section~\ref{SEC:RESULTS}, we shall examine this case, which is a qualitatively different scenario to that considered in \cite{Morrissey:2009ur}, which considered gauge mediation. In the case of sequestering, we shall assume the gravitino to be much heavier than the hidden sector, but, importantly, that the singlet mass-squared $m_S^2>0$ and the hidden gaugino mass-squared $M_\lambda^2$ are of a similar order of magnitude to the hidden Higgs soft terms $m_+^2, m_-^2$, while the hidden $A_S$ term remains small. This is in contrast to gauge mediation where $m_S^2 \sim M_\lambda^2 \sim 0$.

\subsection{Dark matter candidates}
\label{SEC:DMCANDIDATES}

The model above contains essentially two different dark matter
candidates: a Majorana fermion and a Dirac one.\footnote{We are
ignoring the possibility of scalar dark matter since, although
the model as we have written it contains stable scalars, we
expect the symmetries protecting them to be broken at some
higher order in the potential allowing them to ultimately
decay.} Neglecting the effect of kinetic mixing with the
visible neutralino, the fermion mass matrix in the basis
$(\tilde{\lambda}, \tilde{h}_+, \tilde{h}_-, \tilde{s} )$
corresponding to hidden gaugino, hidden Higgsinos and hidden
singlino is given by
\begin{equation} \C{M}_f = \left(
\begin{array}{cccc} M_\lambda & m_{\gamma^\prime} & 0 & 0 \\
m_{\gamma^\prime} & 0 & 0 & 0 \\ 0 & 0 & 0 & \Delta \\
0&0&\Delta & 0 \end{array} \right).
\end{equation}
The Majorana particle is formed from diagonalising the
$\tilde{\lambda}, \tilde{h}_+$ states; in the case of a large
$M_\lambda$, this leads to a see-saw effect and a low mass. We
shall refer to this state as ``$\tilde{o}_1$'', micrOMEGAs
notation for the lightest odd particle. Clearly, there will
therefore always be a fermion lighter than the hidden gauge
boson (to avoid this fate, we would need to add a mass for the
hidden singlino). In order for the Dirac fermion formed from
$\tilde{h}_-, \tilde{s}$ to be the lightest state, we would need
$\lambda_S < \sqrt{2} g_h$ and for the Majorana mass
$M_\lambda$ to be rather small at the high-energy scale (this
could happen, for example, in a string model where the modulus
corresponding to the gauge coupling does not obtain an $F$-term),
although it is somewhat suppressed in running down to the low
scale. Hence, the Dirac fermion scenario is not compatible with
radiative-breaking models, but presents an attractive candidate
for the visible sector induced breaking. We shall refer to this
state as~``$\tilde{o}_7$''. Note that this would not be a good candidate
in gauge mediation, as there the singlet scalar would necessarily be
lighter than the fermion \cite{Morrissey:2009ur}.

In a complete analysis including the couplings and annihilation cross sections, it is necessary to take the mixing with the visible neutralino into account; this we do in appendix~\ref{APP:FERMIONS}.

Finally, we comment on the (lack of) effect of breaking the
residual global symmetry on the above analysis. This could
occur via terms in the super- or K\"ahler potential of the form
$S^n$ suppressed by an appropriate power of a mass scale, such
as the string or Planck scale; for example, in string theory, it
would be natural to expect terms of the form $S^n e^{-aT}$
where $T$ is some modulus charged under a (broken) gauge
symmetry from which the residual global symmetry descends -- the
effect could thus be \emph{exponentially} suppressed by the
expectation value of $T$, and so can, in principle, be naturally
arbitrarily small. Since these are small effects, they will not
affect the hidden gauge symmetry breaking (the singlet field
would obtain a very small expectation value due to the
radiative generation of a tadpole term in the potential, which
would no longer be prohibited by the symmetry, but of course
could be made arbitrarily small), but they will split the Dirac
fermion into two Majorana ones with a potentially undetectable
mass splitting. However, the lightest of these states, when it
is the LSP, will be protected from decay by R-parity. This is
important when considering the constraints of Big Bang
Nucleosynthesis (BBN, and is in contrast to the cases
considered in, e.g., \cite{Katz:2009qq}): in principle, any
unstable relic with a lifetime greater than $\mathcal{O}(100)$
seconds must obey strict constraints on its density during BBN;
see, e.g., \cite{Cyburt:2002uv}. On the other hand, the model
does possess heavy scalars whose decays are protected by this
symmetry, and also the heavier component of the Dirac fermion
would then decay; however, since the effect can be arbitrarily
small, we may simply assume that the lifetimes are many times
that of the Universe, and so we can to all intents and purposes
treat the symmetry as exact. This is our favoured perspective,
but we can alternatively make the breaking strong enough that
the scalars and heavier components can decay fast enough; for
example, a coupling of the form $W\supset \lambda S^3$ will
induce decays of $S$ with $\Gamma \sim 10^{-2}\lambda^2 m_S$,
so $\lambda \gtrsim 10^{-11}$ would suffice; similarly, a mass
splitting of the fermions of $\Delta m_{\Delta}^2$ will allow
decays with $\Gamma \sim 10^{-2} |g_h \chi|^2  m_\Delta
\frac{\Delta m_{\Delta}^2}{m_{\Delta}^2}$ which, for the values
of the couplings considered in this paper, will suffice if
$\frac{\Delta m_{\Delta}^2}{m_{\Delta}^2} \gtrsim 10^{-11} $.
%A last comment on BBN constraints is that, since we are considering
%dark matter candidates in the several GeV range, they are decoupled
%well before it begins and so should not interfere.
We will comment more upon BBN constraints in section \ref{SEC:CONSTRAINTS:BBN}.

\section{Constraints and discovery potential}
\label{SEC:CONSTRAINTS}

There are already a wealth of constraints on the parameter space of models with dark forces and hidden matter that we must apply in our search over models. However, there are also future experiments which will have the potential to rule further regions out -- or make a discovery. In this section, we summarise these current and future constraints and illustrate them by application to a toy model.

\subsection{Limits on the hidden photon}
\label{SEC:CONSTRAINTS:HP}
%\marginnote{\hspace{-0.2cm}1st~sentence \\ added, \\\hspace{-0.2cm}2nd~slightly \\ changed}%
A summary of various constraints on hidden photons from
cosmology (including BBN), astrophysics and laboratory searches
for the whole mass and kinetic mixing ranges $10^{-9} \GeV \leq
m_{\gamma'} \leq 10^{3} \GeV$ and
\mbox{$10^{-15}\leq |\chi|\leq 1$} has been presented for example
in~\cite{Jaeckel:2010ni} and references therein. For the mass
range of interest in this work, the constraints from
electroweak precision tests (EWPT) are used as have been
presented in~\cite{Hook:2010tw}, where the strongest constraint
is provided by the mass of the $Z$ for most of the parameter
space. In the following plots (figures \ref{Fig:ToyModScatt},\ref{RadScanNoSIMPLE},\ref{RadScanWithSIMPLE},\ref{VisibleShowKappa} and \ref{VisibleShowHalos}) of $\chi$ vs $m_{\gamma'}$, this   %\marginnote{fig4-6,8,9} %
is shown as a long-dashed approximately horizontal blue line
excluding roughly $\chi \gtrsim 3 \times 10^{-2}$. Another
constraint comes from the muon anomalous magnetic
moment~\cite{Pospelov:2008zw} and is dominant for $m_{\gamma'}
<$ 1 GeV: in the above-mentioned  plots of $\chi$ vs $m_{\gamma'}$,      % \marginnote{fig4-6,8,9} %
this is a dashed-dotted brown line at low masses and $\chi
> 10^{-2}$. There is also a model-dependent
constraint from BaBar searches~\cite{Hook:2010tw} that might be
the most constraining in the region $0.2 \GeV \lesssim
m_{\gamma'} \lesssim 10 \GeV$ but only applies if the $\gamma'$
can not decay into hidden sector particles; in the above-mentioned
plots of $\chi$ vs $m_{\gamma'}$, this is a dashed dark purple      %\marginnote{fig4-6,8,9} %
line at low masses below 10 GeV and $\chi \sim 2 \times
10^{-3}$. This constraint does apply for most of the
supersymmetric models we are considering, where the mass of the
$\gamma'$ and hidden matter are similar -- preventing a decay of
$\gamma'$ to the hidden sector. However, if the hidden photon
can decay to hidden matter, then there is instead a much weaker
constraint from the $Z$ width; we require
\begin{align}
\frac{\Gamma (Z \rightarrow \mathrm{hidden})}{\Gamma ( Z \rightarrow \nu \ov{\nu})} \lesssim 0.008
\end{align}
which for a single hidden Dirac fermion of mass $M_X < M_Z$ and
unit charge under the hidden \U1 corresponds to (see also
\cite{Essig:2009nc})

\begin{align}
8 c_W^2 s_W^2 (\frac{\s_\phi}{c_\epsilon})^2 \left(\frac{g_h^2}{e^2}\right) ( 1 + 2 \frac{M_X^2}{M_Z^2}) \sqrt{1-4 \frac{M_X^2}{M_Z^2}} \lesssim 0.008
\end{align}
where $c_W, s_W$ are the usual cosine and sine of the weak
mixing angle respectively; $s_\phi$ is defined in equation
(\ref{tanphi}). For $M_X \ll M_Z$, this simplifies to $\chi g_h
\lesssim 0.04$. Clearly, for a small number of hidden particles
(and $g_h < 1$), this is a weaker constraint than the
measurement of the $Z$ mass.

For $m_{\gamma'}$ below 1 GeV, there are additional constraints
which are shown as grey areas in figure~\ref{Fig:ToyModPlot}.
The past electron beam dump experiments
E141~\cite{Riordan:1987aw}, E137~\cite{Bjorken:1988as} and
E774~\cite{Bross:1989mp} have been reanalyzed
in~\cite{Bjorken:2009mm} in terms of hidden photons and were
found to place limits on small masses $\lesssim 2 m_\mu$. In
addition, another such limit has been obtained from an electron
beam dump experiment at Orsay~\cite{Davier:1989wz}
in~\cite{Andreas:2011xf}. Recently, two electron fixed target
experiments A1 at MAMI in Mainz~\cite{Merkel:2011ze} and APEX
at JLab~\cite{Abrahamyan:2011gv} started, which are both
searching for hidden photons behind a thin target from
bremsstrahlung off an electron beam and which where already
able to set first new constraints. Another limit arises
in~\cite{Blumlein:2011mv} from the reanalysis of data from a
proton beam dump taken at the U70 accelerator at IHEP
Serpukhov. At the Frascati DA$\phi$NE $\phi$-factory, the
\mbox{KLOE-2} experiment~\cite{Archilli:2011zc} set further
constraints using $e^+e^-$ collisions. However, not only are
there limits on the kinetic mixing for very light hidden
photons, but excitingly there are also dedicated experiments
planned (and partly already running) that can further probe
this parameter space with real discovery potential. There are
two fixed target experiments (A1~\cite{Merkel:2011ze} and MESA)
in Mainz and three (APEX~\cite{Essig:2010xa,Abrahamyan:2011gv},
DarkLight~\cite{Freytsis:2009bh} and HPS~\cite{HPS}) at JLab.
The estimated sensitivities of those experiments are shown in
figure~\ref{Fig:ToyModPlot} for the toy model.

\subsection{Constraints from Big Bang Nucleosynthesis}
\label{SEC:CONSTRAINTS:BBN}

If a model produces too many high-energy photons, they can
dissociate nuclei (such as lithium) and ruin the predictions
from nucleosynthesis. The thresholds for these processes are of
the order of a few MeV, and so photons produced with energies
above this are potentially dangerous. This is typically used to
constrain long-lived decaying particles where a photon is among
the decay products; due to the rapid interactions of the
photons with the plasma, a ``zeroth order'' spectrum of
energies is produced with a cutoff at $m_e^2/(22 T)$ (where
$m_e$ is the electron mass), and so these reactions only
activate for temperatures $T$ \emph{below} $0.01$ MeV,
corresponding to times of the order of $10^4$s. The strongest
constraints are for particles with lifetimes of $10^8$s. In
models with a hidden sector,  it is then natural to wonder
whether visible photons can be produced, for example, by decays
of particles in the hidden sector or the occasional
annihilation of the frozen-out dark matter particles.

For a \emph{massless} hidden photon, hidden sector matter does
acquire a small charge under the visible photon (they become
``millicharges''), in which case the constraints upon their
presence during BBN are summarised in \cite{Jaeckel:2010ni}.
However, since we are considering a massive hidden photon, the
diagonalisation of the physical states is given in equation
(\ref{ReparamGauge}), from which it can be seen that
hidden sector states do not couple to the visible photon
(cf. also~(\ref{EQ:PhysicalCouplings})). Moreover, once a
hidden photon is produced, the physical state does not
oscillate into visible photons\footnote{Recall that equation
(\ref{ReparamGauge}) is valid in a vacuum, and during BBN there
is a small effect due to the thermal mass for the photon $m_P$
in the plasma. Since we must consider temperatures below $0.01$
MeV, below the electron mass, this is given by $m_{P}^2 \simeq
4\pi \alpha \frac{n_e}{m_e} \simeq 4\pi \alpha
\frac{n_\gamma}{m_e}\eta$, where $n_e, n_\gamma$ are the
densities of electrons and photons, respectively, and $\eta$ is
the baryon-to-photon ratio. For $T = 0.01$ MeV, $\eta =
10^{-9}$, this gives an upper bound on the mass of $m_P \lesssim
10^{-8}$ MeV. The effect of this additional tiny mass is a
minuscule orthogonal rotation of the physical states, whereby
the photon and hidden photon mix by an amount $\chi c_W
\frac{m_P^2}{m_{\gamma'}^2}$. If there were a relic population
of hidden photons, in principle, a tiny fraction of them could
oscillate into visible photons, and we would need to consider
their effect on BBN -- but, further, for the range of hidden
photon masses and kinetic mixing we are considering here, this
is clearly completely negligible.}
(so the constraints will be very different from, for example, possible sterile neutrinos). 
 It does, however, couple to
visible sector matter and decays with a width of $\Gamma
\simeq \frac{1}{3} Q^2\alpha\chi^2 c_W^2 m_{\gamma^\prime} $
into each light species of charge $Q$, i.e. $\Gamma > 10^{-2}
\chi^2 $GeV, or a lifetime $\tau_{\gamma'} <
\left(\frac{10^{-11}}{\chi}\right)^2
\left(\frac{\mathrm{GeV}}{m_{\gamma'}}\right)\rm{s}$. In this
work, we shall be considering $\chi > 10^{-5}$, for which the
hidden photon will always decay immediately on any cosmological
timescales -- and so there will be no relic density of hidden
photons present.

From the above, we can see that BBN constraints will not affect
our dark matter models in much the same way that they do not
restrict standard weakly interacting massive particles (WIMPs). However, to be completely strict, let
us consider that the annihilation of our dark matter particle
will have some non-zero but small branching ratio into
visible-sector photons, which we denote $r_\gamma$. One could
imagine that this would arise from the plasma-induced mixing
described above, where $r_\gamma \sim \chi c_W
\frac{m_P^2}{m_{\gamma'}^2}$, but given the parameter region we
are considering, this will be dominated by loop effects
instead.
%via the parity-violating terms (see equation \ref{EQ:PhysicalCouplings}) -
%i.e. $r_\gamma \sim \alpha \alpha_h \frac{m_{\gamma'}^4}{m_Z^4}$.
%Since we shall consider $m_{\gamma'}< 20$ GeV, we have $r_\gamma \lesssim 10^{-6}$.
Since the hidden $U(1)$ is not anomalous, the first diagram
appears at two loops, yielding $r_\gamma <
\frac{\alpha^2}{(4\pi)^2} < 10^{-6}$.

The rate of annihilations of our dark matter candidate $\psi$
into photons \emph{per unit volume} (assuming that it annihilates entirely through the
hidden photon channel) is $\Gamma_\gamma/V = r_\gamma n_\psi^2
\bra \sigma v \ket$, where $n_\psi$ is the relic density. The
strongest bounds for BBN arise for particles of lifetime
$10^8$s and constrain \cite{Cyburt:2002uv}:
\begin{align}
m_\psi \frac{n_\psi}{n_\gamma} < 5.0 \times 10^{-12} \ \mathrm{GeV}.
\end{align}
We can therefore take a rough constraint by requiring that our
relic particles never produce more photons than such a decaying
particle; i.e. $\Gamma_\gamma/V < \frac{n_\gamma}{m_\psi} \times
5.0 \times 10^{-12} \ \mathrm{GeV}/10^8 \rm{s} $ for
temperatures lower than $0.01$ MeV. This yields, roughly,
\begin{align}
r_\gamma \lesssim 2 \times 10^{-3} \ \left( \frac{0.01 \ \mathrm{MeV}}{T_c} \right)^3 \left(\frac{T_f}{50 \ \mathrm{MeV}}\right)
\end{align}
where $T_f$ is the freezeout temperature (typically $T_f \sim
m_\psi/20$) and we compare the rates at temperature $T_c< 0.01$ MeV. This is an overly conservative bound (since the largest disruptive effect of a decaying particle occurs at temperatures much below $0.01$ MeV) but even so is very weak  %(indeed,
%it is weaker than the constraints due to self-interactions of
%dark matter)
and will not affect the rest of our analysis.

%we must recall that the above is valid in a vacuum, and during
%BBN there is a small effect due to the thermal mass for the
%photon $m_P$ in the plasma. Since we must consider temperatures
%below $0.01$ MeV, below the electron mass, this is given by
%$m_{P}^2 \simeq 4\pi \alpha \frac{n_e}{m_e} \simeq 4\pi \alpha
%\frac{n_\gamma}{m_e}\eta$, where $n_e, n_\gamma$ are the
%densities of electrons and photons respectively, and $\eta$ is
%the baryon-to-photon ratio. The effect of this additional small
%mass - since we are in the regime $m_P \ll m_{\gamma'}$ - is a
%tiny orthogonal rotation of the physical states, whereby the
%photon and hidden photon mix by an amount $\chi c_W
%\frac{m_P^2}{m_{\gamma'}^2}$. Hence if hidden photons are
%produced by occasional annihilations by frozen-out particles,
%then a tiny number of visible photons can also be produced. To
%quantify this, the rate of these annihilations will be

\subsection{Limits from dark matter}
\label{SEC:CONSTRAINTS:DM}

There are further experimental constraints arising on the dark
matter particle, its mass and its interactions. First of all,
the dark matter particle should not have a relic abundance in
excess of the one measured by WMAP~\cite{Komatsu:2010fb},
\begin{equation}
\Omega_{\rm DM} h^2 = 0.1123 \pm 0.0035.
\label{EQ:WMAP7}
\end{equation}
This is a very strict limit and translates to a lower limit on
the dark matter (DM) annihilation cross section. We compute the
dark matter relic abundance using micrOMEGAs where we have
implemented our model. However, while there is an upper limit
on the relic abundance, there is no objection to having a dark
matter candidate whose abundance is lower than the one
measured. In this case, it would then only be a part of the
total dark matter (we shall refer to this as subdominant DM),
and the remaining dark matter density would consist of other
particle(s) such as an axion or axion-like particle whose
phenomenology is not the subject of this article -- we shall
simply assume in such cases that the direct detection cross
sections and interactions with the hidden sector of the
additional dark matter are both negligible. In all of our plots, we     %  \marginnote{all fig} %
show parameter points that give an abundance in agreement with
the WMAP value in dark green and ones where the DM is
subdominant in light green.

Additional constraints apply to the dark matter particle and
its scattering cross section on nuclei. It is necessary to
distinguish spin-dependent (SD) and spin-independent (SI)
scattering. Depending on whether the dark matter particle is a
Majorana or Dirac fermion, it has either dominantly SD or SI
interactions, respectively. The SI interaction is, moreover, dominated by
$\gamma'$ exchange, which couples almost exclusively to the proton, particularly at
low hidden photon masses (where the mixing can be treated as being effectively between the photon and hidden photon -- see appendix \ref{APP:KM}). The SI interaction is therefore strongly isospin-dependent, and we must rescale limits on the cross sections accordingly (which usually assume equal couplings for protons and neutrons). For the SD interactions, however, the isospin dependence is rather weak, being dominated by $Z$ exchange.
Current limits from direct dark
matter detection experiments are strongest for SI scattering
cross sections ($\sim 10^{-42} \sqcm$), while SD cross sections both on protons and on neutrons
only start to be excluded at the $10^{-38} \sqcm$ level.

On the SI side, for the low dark matter masses ($\sim 10 \GeV$)
we are interested in, the most relevant constraints come from
XENON and CDMS. However, due to the signal claims from DAMA and
CoGeNT,\footnote{We have not explicitly included the CRESST
signal in our search. One of their two signal regions is
roughly compatible with both DAMA and CoGeNT signals, although
this is still subject to astrophysical uncertainties.} there
has been a large debate on the reliability of those
constraints, especially at low dark matter masses close to the
energy threshold of the experiments. There are also large
astrophysical (halo model, dark matter velocity and local dark
matter density) and nuclear physics uncertainties that should
be taken into account. Even though XENON and CDMS claim to rule
out most of the DAMA and CoGeNT preferred regions, the positive
signals remain and there have been various studies of how to
reconcile those different results.\footnote{One interesting
possibility is to allow isospin-dependent interactions with
just the right behaviour to suppress the interaction cross
section with xenon
nuclei~\cite{Giuliani:2005my,Feng:2011vu,Frandsen:2011ts,McCabe:2011sr,Frandsen:2011cg,Gao:2011ka,Cline:2011zr}.
We simply note that, although in the case of hidden Dirac
fermions the interaction is almost entirely with protons rather
than neutrons, in our models this tuning is not possible.} We
adapt the analysis of~\cite{Arina:2011si} which made a
systematic scan taking into account the various uncertainties.
There it is found that depending on the halo model, some of the
CoGeNT and sometimes even DAMA preferred region is consistent
with the exclusions from XENON and CDMS. For the details of the
different halo models, see~\cite{Arina:2011si}; we will mostly
use their so-called Standard Model Halo (SMH) and in a few
cases show the differences that arise when changing for example
to a Navarro-Frenk-White (NFW) or an Einasto profile.

We strictly apply the XENON100 and CDMSSi constraints derived
in~\cite{Arina:2011si} to the SI scattering cross sections and
only show points that are not excluded by any of the two     %\marginnote{fig7bot,11} %
experiments. In the plots of $\sigma_p^{\rm SI}$ \footnote{We
always show scattering from protons in the plots, hence
$\sigma_p^{\rm SI}$, since the constraints are strongest for
these, and because our Dirac candidate will couple more
strongly to protons than neutrons.} vs $m_{\rm DM}$ in section
\ref{SEC:RESULTS} (see figures \ref{RadScat} and
\ref{VisibleSI}), the CDMS limit is shown as a dashed turquoise
line, while XENON100 is a dashed-dotted blue line. For most halo
models, CDMS is more constraining at lower masses than XENON100.

In the SD case, there are both for scattering on protons and on neutrons several direct detection
experiments sensitive to the low dark matter masses we are
interested in. Different papers also tried to explain the DAMA
signal by spin-dependent scattering either exclusively from
neutrons~\cite{Ullio:2000bv} or from
protons~\cite{Kappl:2011kz}. The former case is, however, not
applicable in our models, as the spin-dependent cross sections
of the Majorana fermion are always of the same order of
magnitude both for protons and neutrons. In the latter
analysis, it was shown that for scattering on protons, the DAMA
favoured region is ruled out by Super-Kamiokande due to neutrinos from DM
annihilation in the Sun almost independently of the
annihilation channel. Additionally, the cross sections required
in both scenarios are more than one order of magnitude above
the largest ones that can be obtained in our models. Therefore, if the explanation of the DAMA (and CoGeNT) signals
is confirmed as arising from spin-dependent scattering, it would rule out the models considered in this paper.
Hence, we do not study this in more detail and simply apply the
various spin-dependent scattering direct detection constraints.
Until June 2011, PICASSO for the lightest and COUPP for the
slightly larger masses were the most constraining
experiments for SD scattering on protons~\cite{Behnke:2010xt}. Very recently, a new direct
detection experiment SIMPLE~\cite{Felizardo:2011uw} has
published a limit on the SD scattering cross section on protons which in
the low mass range is one order of magnitude stronger than
previous experiments (for a critique of their limit,
see~\cite{Collar:2011kr} and the collaboration's
response~\cite{Collaboration:2011ig}). There is also a quite
strong limit from Super-K using neutrino fluxes produced by
dark matter annihilation in the Sun which, however, only applies
to dark matter masses above $20$ GeV (only neutrino-induced
upward through-going muons have been used in this analysis
which leads to a quite high-energy threshold and therefore a
sensitivity only to larger DM
masses)~\cite{Desai:2004pq}.\footnote{There is another more
recent analysis~\cite{Kappl:2011kz} with limits for smaller
masses. Application of these limits taking into account the
annihilation details and branching ratios is beyond the scope
of this work and left for future works~\cite{inPrep}.} For SD scattering on neutrons there are limits from XENON10~\cite{Angle:2008we}, Zeplin~\cite{Akimov:2011tj} and CDMS~\cite{Ahmed:2010wy,Ahmed:2008eu,Ahmed:2009zw}, the strongest of which, set by XENON10 for the mass range of interest in this paper, is less constraining than the SIMPLE limit. 

In the following analysis, we use all constraints from SD scattering both on protons and on neutrons with the exception of SIMPLE as
strict exclusions and show only points consistent with those
limits. As there has been criticism of SIMPLE's limit, we will
not apply this universally but rather show how our results
change when taking it into account. In the plots of  
$\sigma_p^{\rm SD}$ vs $m_{\rm DM}$ in
section~\ref{SEC:RESULTS} (see figures~\ref{RadScat}
and~\ref{VisibleSDMajorana}), the exclusion lines for the
different experiments are as follows: SIMPLE short-dashed brown
line, Super-K dashed black line, PICASSO long-dashed orange 
line, COUPP2011 dashed-dotted turquoise line, COUPP2007 
dotted blue line and KIMS long-dashed green line. The plots of $\sigma_n^{\rm SD}$ vs $m_{\rm DM}$ in the same figures show the limits of XENON10 as dashed-dotted blue, Zeplin as dotted pink and CDMS as dashed turquoise lines.

Those constraints on the scattering cross section can strictly
only be applied to particles that actually constitute the
entire dark matter density. If the dark matter is subdominant
however, the limits on its scattering cross section have to be
rescaled accordingly: the local density $\rho_\psi$ of a dark
matter candidate  $\psi$  relates to the local total DM density
$\rho_{\rm DM}$ as their abundances
\begin{equation}
\frac{\rho_\psi}{\rho_{\rm DM}} = \frac{\Omega_\psi}{\Omega_{\rm DM}}
\end{equation}
and so do the limits that are set by direct detection (DD)
experiments. Thus, an experimental bound on $\sigma_{\rm DD}$
translates into an actual bound on the scattering cross section
$\sigma_\psi$ of $\psi$ as
\begin{equation}
\sigma_\psi = \sigma_{\rm DD}  \frac{\Omega_\psi}{\Omega_{\rm DM}}.
\end{equation}
This means that direct detection constraints on the scattering
cross section become less potent for subdominant DM
particles.\footnote{This is obviously based on the reasonable
assumption that the local DM has the same content of different
DM contributions as averaged over the whole Universe.}

\subsection{Application to toy model}
\label{SEC:TOY}

To illustrate the above constraints/future experimental reach,
and more importantly provide a comparison to the more complete
model of section \ref{SEC:SUSYDS} that we shall investigate in
section \ref{SEC:RESULTS}, here we shall consider a toy model.
This is the simplest possible dark sector: a Dirac fermion
$\psi$ with unit charge only under the (massive) hidden \U1. We
shall not include any Higgs sector -- the \U1 could, after all,
naturally have a GeV scale mass via the St\"uckelberg mechanism
\cite{Goodsell:2009xc,Cicoli:2011yh} -- so we will not consider
how the dark matter particle becomes massive. This is
essentially the model considered in
\cite{Dudas:2009uq,Mambrini:2009ad,Mambrini:2010yp,Mambrini:2010dq,Mambrini:2011dw}
except that we shall insist on the relation
(\ref{ChiRelation}); the parameters are the dark matter mass
$m_{\psi}$, hidden photon mass $m_{\gamma'} $, kinetic mixing
$\chi$ and the tuning parameter~$\kappa$.

\subsubsection{Constraints and future searches}

\begin{figure}[htb!]
\begin{center}
\includegraphics[width=9cm]{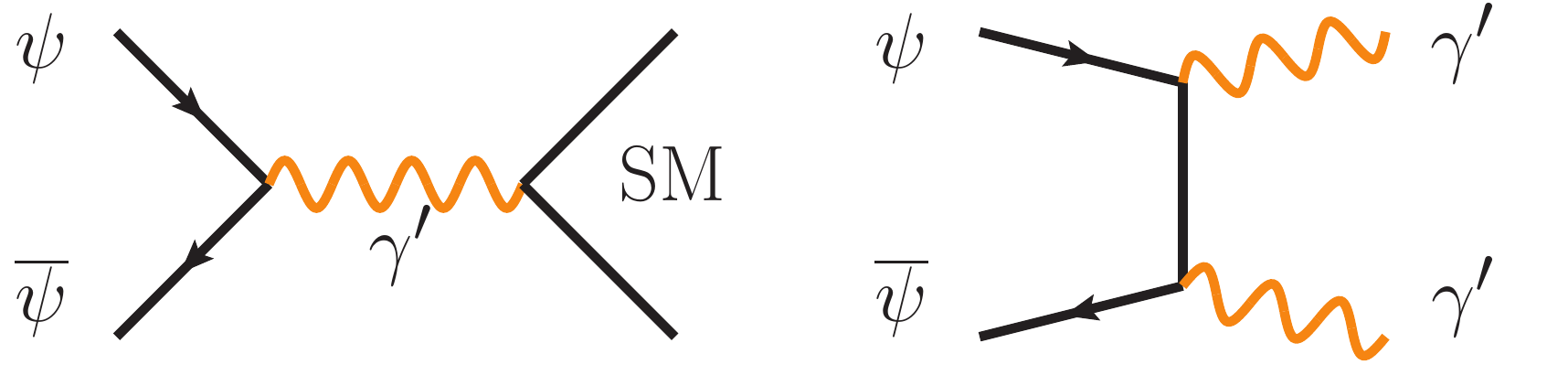}
\caption{Annihilation diagrams: $s$-channel annihilation on the left,
resonant at $m_{\gamma'} =2m_{\psi}$; $t$-channel on the right,
accessible and dominant when $m_{\psi}>m_{\gamma'}$.
\label{Fey:DMAnn}}
\end{center}
\end{figure}

\begin{figure}[htb!]
\begin{center}
\includegraphics[width=0.49\textwidth]{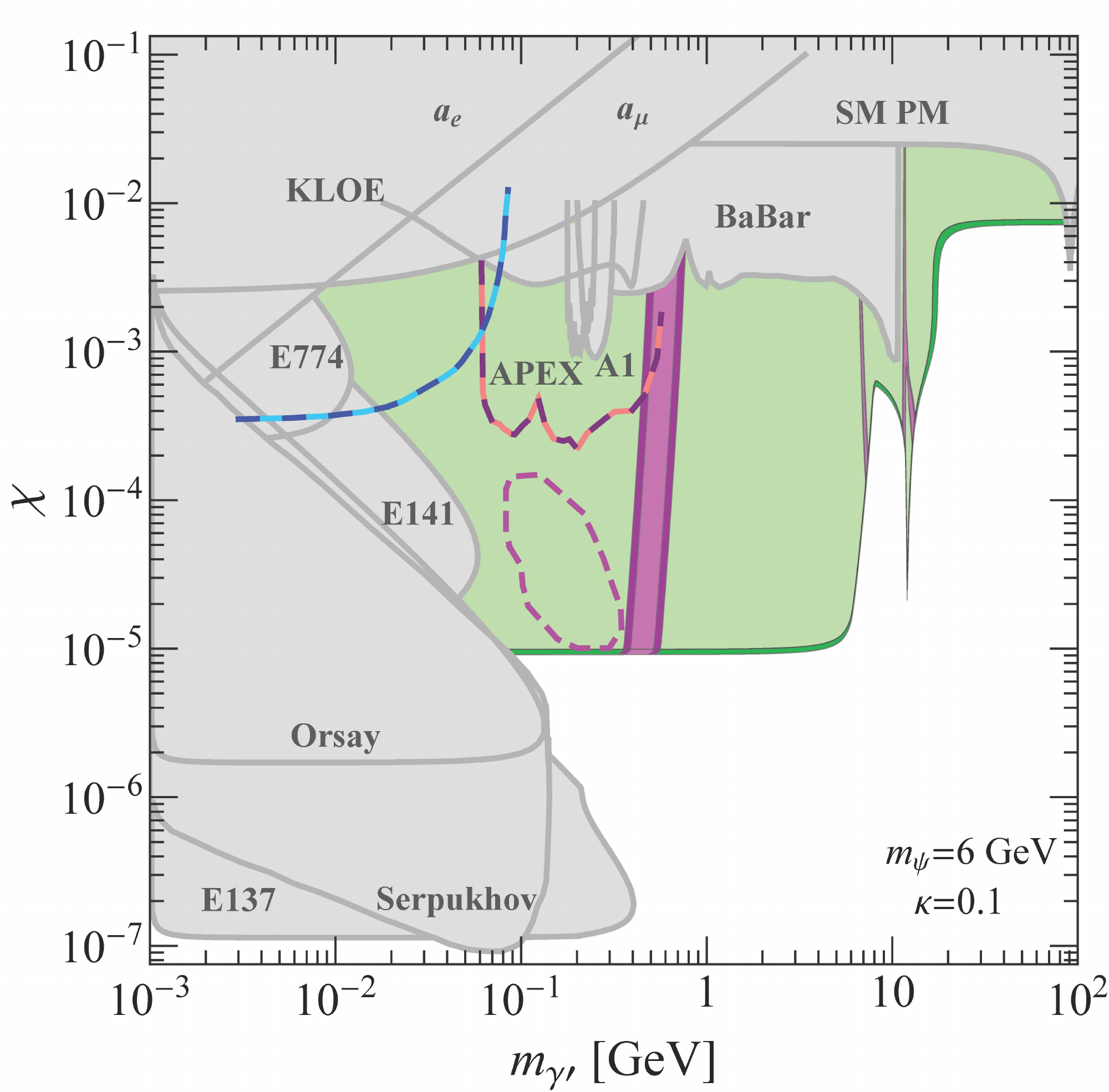}
%\includegraphics[width=7cm]{PlotsToyMod/Plot_chivsmgamma_ToyMod_SMH_resc_mDM6GeV_kappa10_final.pdf}
%\end{center}
%\begin{center}
\includegraphics[width=0.49\textwidth]{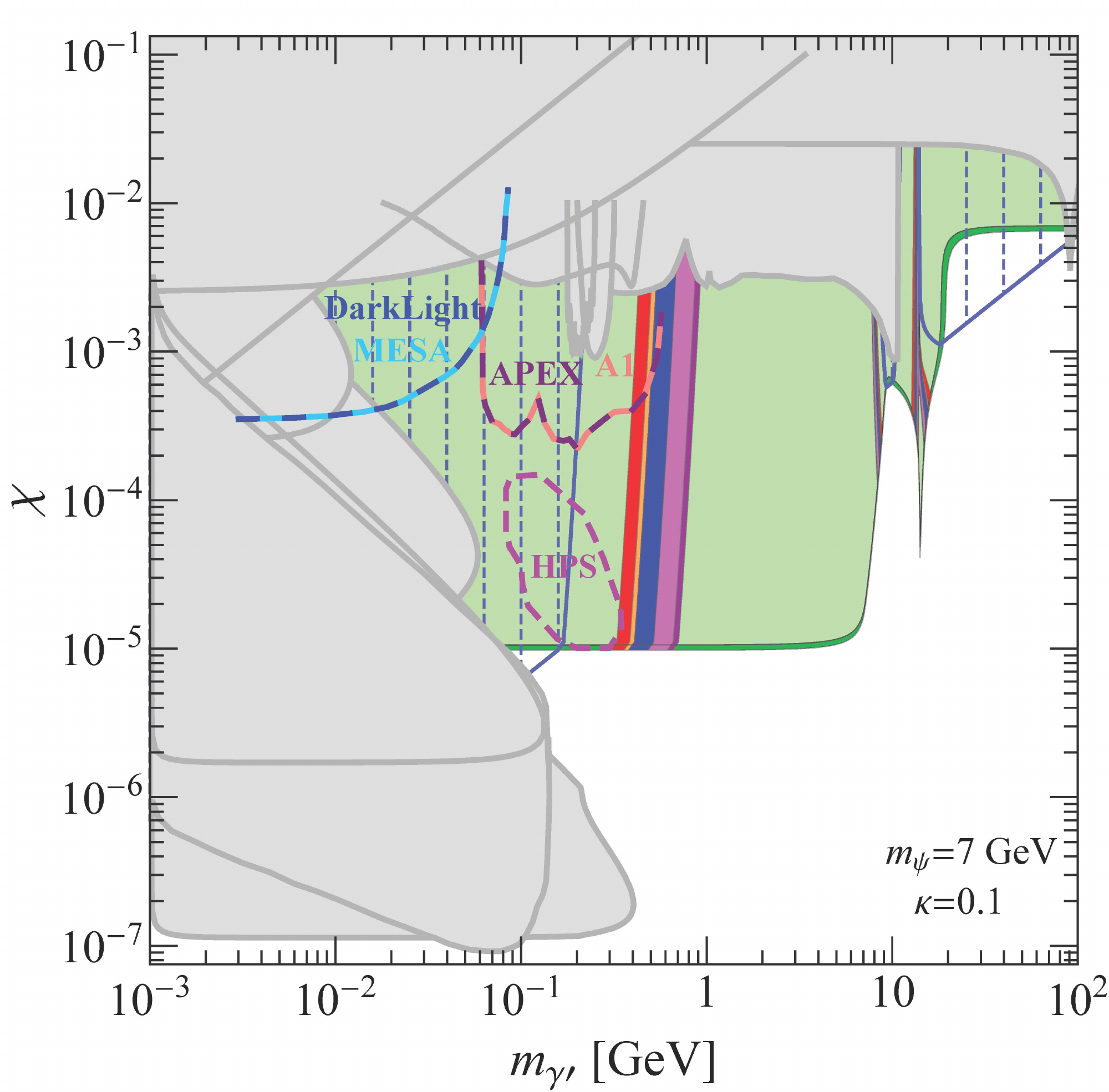}
%\marginnote{\hspace{1.7cm}plot $\kappa=10$ removed}
\caption{Dark matter (DM) relic abundance and direct detection cross
section in agreement with CoGeNT and DAMA for a Dirac fermion
DM candidate with a mass of 6 GeV in the Standard Halo Model
(SMH) (\textit{left}) and 7 GeV in the Einasto halo model
(\textit{right}) with $\kappa=0.1$ for both. The grey regions
are excluded by different searches as shown in the
\textit{left} plot: the anomalous magnetic moment of electron
and muon $a_e$ and $a_\mu$, electroweak precision test EWPT,
model-dependent BaBar searches, $e^+e^-$ collisions in KLOE, the
electron fixed target experiments A1 and APEX, the electron
beam dump experiments E774, E141 E173 and Orsay as well as the
proton beam dump at Serpukhov (cf.
section~\ref{SEC:CONSTRAINTS:HP} for details and references). The coloured
lines with corresponding labels in the \textit{right} plot correspond to sensitivities of
the already running experiments A1 and APEX as well as the
planned fixed target experiment HPS (see section
\ref{SEC:CONSTRAINTS:HP}). The thin dark green regions give the correct
relic abundance, light green is subdominant DM, and white is
overabundant and therefore excluded. The
scattering cross sections are such that they can explain the
CoGeNT observation in the purple area, the DAMA observation
in the red area, and both experiments at the same time in the
blue area (only possible for the Einasto profile in the right
plot). The blue line in the right hand plot
is the XENON100 bound which excludes all the dashed shaded area
above the line. \label{Fig:ToyModPlot}}
\end{center}
\end{figure}

The DM can annihilate through and/or into hidden photons
according to the diagrams shown in figure~\ref{Fey:DMAnn}.
Whereas the left diagram is possible for all DM masses, the
right one is kinematically only accessible when $m_{\psi} \geq
m_{\gamma'}$. The left diagram also leads to a resonant
enhancement of the annihilation cross section and accordingly
to a dip in the relic abundance for $m_{\gamma'} = 2 m_{\psi}$.
This can been seen in figure~\ref{Fig:ToyModPlot} where we show
the relic abundance for a dark matter mass of 6 GeV (left plot)
and 7 GeV (right) as a function of the kinetic mixing $\chi$
and the hidden photon mass $m_{\gamma'}$. The grey areas are
excluded by beam dump experiments (curves on the left side of
the plot), muon and electron anomalous magnetic moment (top
left corner of the plot) as well as the BaBar search and EWPT
(curves at large $\chi$ and $m_{\gamma'}$) as described in
section~\ref{SEC:CONSTRAINTS:HP}. The thin dark green band is
the region which gives the correct WMAP abundance
(\ref{EQ:WMAP7}), while in the light green areas, the DM
candidate is subdominant. The white region is excluded since it
gives a too large relic abundance. For very small hidden photon
masses, annihilation proceeds only via the left diagram of
figure~\ref{Fey:DMAnn} and is essentially independent of
$m_{\gamma'}$. Therefore, the relic abundance is given by the
kinetic mixing only, which itself is determined by the hidden
gauge coupling up to a factor $\kappa$.\footnote{We have
investigated varying $\kappa$ within the an order of magnitude,
and it does not make a qualitative difference to the plots.}
%is shown in figure~\ref{Fig:ToyModPlot} where the two
%extreme cases of its range are used in the two upper plots.
The coloured lines with named labels represent the future searches
mentioned in section~\ref{SEC:CONSTRAINTS:HP}, which as can be
seen from the plot will probe portions of the interesting
parameter space.

The DM particle considered for the toy model in this section
can also scatter elastically on nuclei. As it is a Dirac
fermion, this process is spin-independent, and the corresponding
cross sections can be compared to the positive observations of
DAMA and CoGeNT. The results in figure~\ref{Fig:ToyModPlot} are
given for the Standard Halo Model (SMH) (left plot) and Einasto (right plot).
In the former case, a part of the CoGeNT allowed region -- and in the latter case, both of the CoGeNT and DAMA allowed regions -- are not excluded by the other experiments (CDMS and XENON100). The band in purple/red corresponds to the 90\% (lighter) and 99\% (darker) contours where the correct cross section for CoGeNT/DAMA can be obtained, respectively. The blue band on the right plot gives the region where both DAMA and CoGeNT can be explained at the same time. At the place
where these bands overlap with the dark green region, the DM
candidate that fits the respective DD experiment is also providing all of the dark
matter in the Universe. In the larger part, however, where the
coloured bands are on top of the light green area, the DM particle
explains the corresponding DD signal while only contributing
subdominantly to the total DM.

Constraints on SI scattering from CDMS and XENON100 do not apply to the low DM mass of 6 GeV used in the 
left plot of figure~\ref{Fig:ToyModPlot}. In the right plot, however, for a DM mass of 7 GeV  in the Einasto profile, the scattering cross section is constrained by XENON100 (below the reach of CDMS) which is shown as a blue 
line excluding all the parameter space above it (where there are dashed vertical lines). Where 
the XENON100 exclusion bound enters the WMAP allowed (light green) region, the limit is rescaled as 
described above to correspond to the appropriate dark matter density. 
However, outside of this
region, it is not rescaled -- the straight line shown corresponds
to the behaviour for a constant dark matter density equal to
that observed. This accounts for the sudden change in gradient.
Note that relation (\ref{ChiRelation}) has a significant effect
upon the behaviour of this bound. Outside of the WMAP allowed region, i.e. when we are applying the XENON100 bound for a fixed dark matter density, the corresponding cross section follows a
contour of $\chi \propto m_{\gamma'}$, rather than $\chi
\propto m_{\gamma'}^2$ which we would find if we were instead
keeping $g_h$ constant. This arises since the cross section
behaves as
\begin{align}
\sigma_{\rm DD} \propto& \frac{\chi^2 g_h^2}{m_{\gamma'}^4} \propto \frac{\chi^4}{m_{\gamma'}^4}.
\end{align}
This explains the straight line portion of the XENON100 bound in 
the $\log-\log$ plots of figure~\ref{Fig:ToyModPlot}, where $\kappa$ is held fixed. Note how
this changes when we take rescaling into account: since the
thermal-averaged $\psi$-$\ov{\psi}$ annihilation cross section
multiplied by speed $\bra \sigma_{\rm Ann} v \ket$ for fixed
dark matter and hidden photon mass is proportional to either
$g_h^2 \chi^2$ or $g_h^4$, which according to equation
(\ref{ChiRelation}) translates into $\bra \sigma_{\rm Ann} v
\ket \propto \chi^4$, and as the relic density is proportional
to $1/\bra \sigma_{\rm Ann} v \ket $, we find
\begin{align}
\sigma_{\psi} \propto& \frac{\chi^4}{m_{\gamma'}^4} \frac{1}{\bra \sigma_{\rm Ann} v \ket} \propto \frac{1}{m_{\gamma'}^4}.
\end{align}
Hence the rescaled XENON100 exclusion bound is approximately a
vertical line on the plot, as can be seen in figure
\ref{Fig:ToyModPlot} where the blue line meets the green
band.

\subsubsection{Example data point}

To illustrate the above model, let us consider an example set
of values that satisfies all of the constraints and explains the
signals while constituting all of the dark matter. Taking, as
in figure~\ref{Fig:ToyModPlot}, $\kappa = 0.1$ and a  hidden
dark matter particle of $6$ GeV, we find $\chi=1.2 \times
10^{-5}$, giving via equation~(\ref{ChiRelation}) a hidden gauge
coupling of $g_h = 0.053$; and masses for the hidden photon
between $0.26$ and $0.33$ GeV with width between $2.5 \times
10^{-13} $ and $3.5 \times 10^{-13}$ GeV for the given mass
range. This yields the dark matter density within three
standard deviations of the WMAP7 result (\ref{EQ:WMAP7}), while
the \emph{rescaled} direct detection nucleon cross sections
range from $2.7 \times 10^{-40} \cm^2$ for the smaller hidden
photon mass and  $1.1 \times 10^{-40} \cm^2$ for the higher,
which can explain the CoGeNT signal with the standard halo
model, or both DAMA and CoGeNT when Einasto is used~\cite{Arina:2011si} -- the interaction is almost entirely spin-independent and with the proton. Since the dark matter is so
much heavier than the hidden photon here, it annihilates almost
entirely via the $t$-channel diagram of figure \ref{Fey:DMAnn}
which is unsuppressed by the kinetic mixing relative to the
first; this is then almost independent of the hidden photon
mass; hence, the contour in figure~\ref{Fig:ToyModPlot} of dark
matter density matching the observed one is approximately
horizontal up to masses near that of the dark matter particle.

\subsubsection{Scanning over dark matter masses}

\begin{figure}[htb!]
\begin{center}
\includegraphics[width=0.49\textwidth]{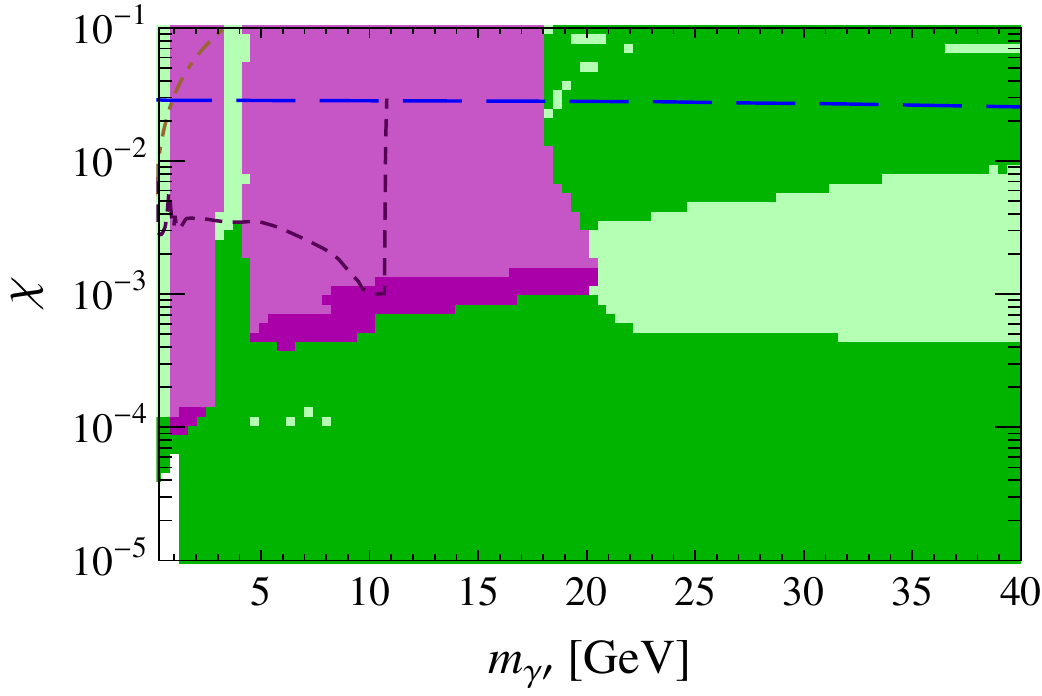}
\includegraphics[width=0.49\textwidth]{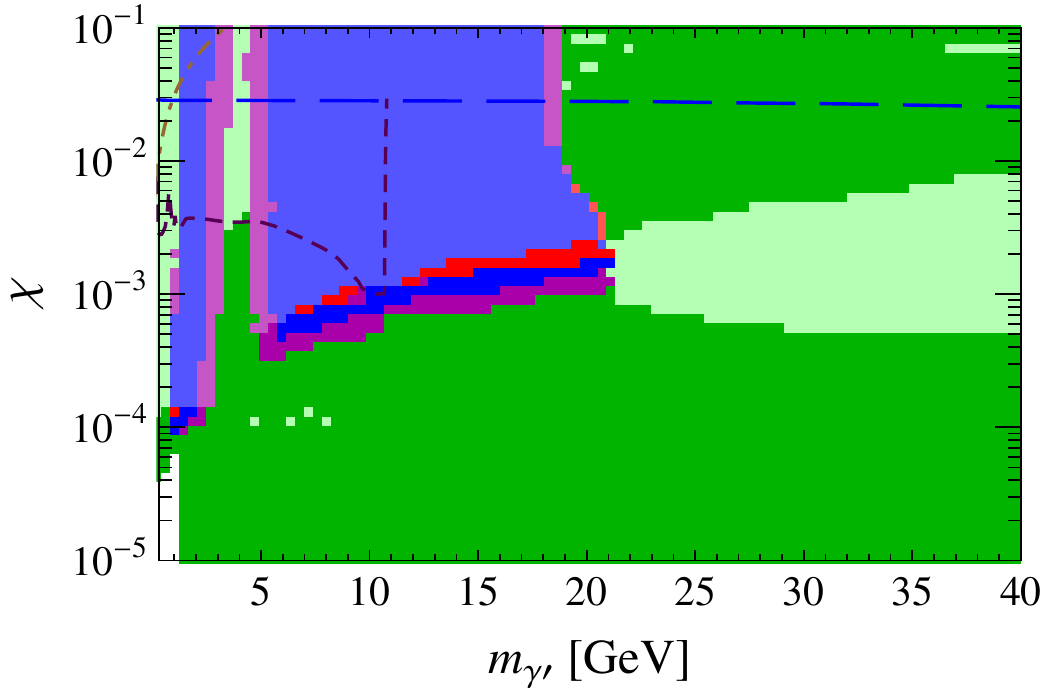}
%\marginnote{\hspace{1.7cm}plot NFW removed}
%\end{center}
%\begin{center}
%\includegraphics[width=7cm]{PlotsToyMod/ScatPlot_chivsmgamma_ToyMod_NFW_resc_kappa1_final.pdf}
\caption{Scan over hidden photon mass $m_{\gamma'}$, kinetic mixing $\chi$ and hidden Dirac
fermion mass $m_\psi$ over the range from 0.8 GeV to 25 GeV, using
(\emph{left}) the Standard Halo Model and (\emph{right}) Einasto
constraints given in \cite{Arina:2011si}. Dark coloured regions
indicate that the correct relic abundance can be found, and lighter
colours indicate that the hidden fermion is a  subdominant dark
matter candidate. Green regions are thus simply WMAP allowed,
but also shown are regions where the direct detection cross
section can explain the signal from either CoGeNT (purple),
DAMA (red) or both at the same time (blue). The parameter
$\kappa$ has been fixed to 1.  For subdominant DM, the
scattering cross sections have been rescaled. All points shown
in the figure are in agreement with all direct detection
constraints. The constraint from electroweak precision tests is
shown as the almost horizontal long-dashed blue line, the
(model-dependent) BaBar limit is shown as a dashed dark line,
and the muon $g-2$ constraint is given as dashed-dotted line at
the top left corner of the plot. \label{Fig:ToyModScatt}}
%\marginnote{\hspace{1.7cm}need~to adjust~text and~caption}
\end{center}
\end{figure}

To fully examine the parameter space of these models, we
performed a scan over the mass $m_\psi$ of the particle $\psi$
for values between 0.8 and 25 GeV while also varying the
kinetic mixing and hidden photon mass. The resulting scatter
plots are shown in figure~\ref{Fig:ToyModScatt} for two
different halo models (for details of the halo models,
see~\cite{Arina:2011si}). For these plots, the
parameter~$\kappa$ was fixed to its central value of one. The
colouring is as follows: dark shades correspond to the dark
matter candidate producing the observed relic abundance, and
lighter shades indicate it is subdominant; green regions do not
correspond to an experimental signal but are not excluded;
purple corresponds to explaining the CoGeNT signal; the DAMA
signal is explained in the red regions; CoGeNT and DAMA signals
are explained simultaneously in the blue regions.  For the
Standard Halo Model (SMH), the CoGeNT and DAMA signal regions
do not overlap, and this is reflected in the absence of blue in
our SMH plots. However, for the Einasto (figure \ref{Fig:ToyModScatt}, right) halo model, %\marginnote{fig4l,8?}%
there is a small region of overlap in mass--cross section space
for the signal regions, which translates into blue regions of
our plots with that choice of halo (figures \ref{Fig:ToyModScatt}, right, and \ref{VisibleShowHalos}, right -- we have also checked that  %\marginnote{fig4r,9r?}%
the situation is very similar for the NFW halo model). However,
in the following section, we shall use mostly the standard halo
model; the choice of halo has a more dramatic effect on the
presence (or otherwise) of overlap of the signal region than
the allowed parameter space of our models.

In the previous subsection, we found that models with very low
mass hidden photons ($<$ GeV) coupled to a Dirac fermion of
mass of a few GeV can be consistent with all constraints, can
form the entirety of the dark matter and can explain the direct
detection signals. As can be seen from the right plot in
figure~\ref{Fig:ToyModScatt}, this is also possible for a thin
(red) band of the parameter space at higher masses. Since the
direct detection signals are only explained for a dark matter
particle in a narrow range of masses between $5.5$ and $8.9$
GeV, we see that this band begins at hidden photon masses equal
to the dark matter mass (indeed  this data is \emph{almost}
enough to read off the parameters of the models from the
scatter plot). This means that the dark matter annihilation
only proceeds via the $s$-channel exchange of
figure~\ref{Fey:DMAnn}; for hidden photons lighter than this,
the $t$-channel annihilation is resonant, explaining the
pole-like shape of the purple patch. A sample model
constituting the entirety of the dark matter, obeying all
constraints and explaining DAMA and CoGeNT (when Einasto is
used) with $m_\psi = 6$ GeV and a spin independent nucleon
cross section of $1.1 \times 10^{-40} \cm^2$, has $\kappa=1$,
$\chi=0.0016$ (thus, $g_h = 0.72$) and $m_{\gamma^\prime} =
14.1$ GeV. The hidden photon  is then quite wide: it has width
$0.14$ GeV, almost entirely decaying into the dark matter.

\section{Analysis of a supersymmetric dark sector}
\label{SEC:RESULTS}

In this section, we describe the results of a scan over the
parameter space of the model of section \ref{SEC:SUSYDS},
constrained by dark matter abundance and direct detection cross
sections. This was achieved by implementation of our models in
micrOMEGAs
\cite{Belanger:2001fz,Belanger:2006is,Belanger:2007zz,Belanger:2008sj,Belanger:2010pz} which automatically computes all of the required annihilation cross sections and integrates the Boltzmann equations to give the relic density. It also calculates the direct detection cross sections for protons and nucleons. Generation of the model files was performed using LanHEP
\cite{Semenov:1996es,Semenov:1998eb,Semenov:2002jw,Semenov:2008jy,Semenov:2010qt}.
We included all of the interactions between the hidden and
visible sector including the neutralino mixing (described in
appendix~\ref{APP:FERMIONS}) and Higgs portal term
(described in appendix \ref{APP:HIGGS}) which we believe to be
novel results; as a result, there is some dependence on the
visible sector spectrum and couplings. Since we were
investigating the effects of gravity mediation, and for
minimality, we chose the visible sector to consist of the MSSM
with a Higgs mass above the LEP bound and the lightest visible
sector neutralino in the range $100$ to $200$ GeV; the effect
of changing the spectrum within these ranges leads to
quantitative changes of a few percent, but not qualitative
ones.

As mentioned in section \ref{SEC:SUSYDS}, we shall take the
kinetic mixing parameter $\chi < 0 $ so that the field $H_+$
obtains a vacuum expectation value (vev) rather than $H_-$. Due to the symmetry of the
model, this is entirely a matter of choice, so the physical
results are unchanged by changing the sign. Therefore, and for
ease of comparison with the previous section, in our plots \ref{Fig:ToyModPlot}--\ref{RadScanWithSIMPLE}, \ref{VisibleShowKappa} and \ref{VisibleShowHalos}, we %\marginnote{fig3-6,8,9} %
show the magnitude of $\chi$.

\subsection{Radiative breaking domination}\label{SEC:RADBREAK}

\subsubsection{Parameter scan}

Here we perform a scan over $\lambda_S$, $\chi$ and
$m_{\gamma'}$ in order to find parameter combinations which
give a light dark matter candidate (mass in the range between
0.8 and 20 GeV) which, as mentioned in section
\ref{SEC:DMCANDIDATES}, we find to be exclusively a Majorana
fermion $\tilde{o}_1$. We insist that $\lambda_S$ and the
hidden gauge coupling inferred from $\chi$ remain perturbative;
this places an upper limit upon $\chi$ via
equation~(\ref{ChiRelation}). We are interested in light hidden
gauge bosons, so we choose a maximum value of $m_{\gamma'}$ of
$40$ GeV.  The low-energy parameters are found by choosing
boundary conditions at the high-energy scale ($10^{16}$ GeV)
and running down; this ensures that we have \textit{bona fide}
consistent models at the low-energy scale, rather than choosing
the parameters completely \textit{ad hoc}. This search uses the
RGE engine from SoftSUSY \cite{Allanach:2001kg}.

\begin{figure}[p]
\begin{center}
\includegraphics[width=0.49\textwidth]{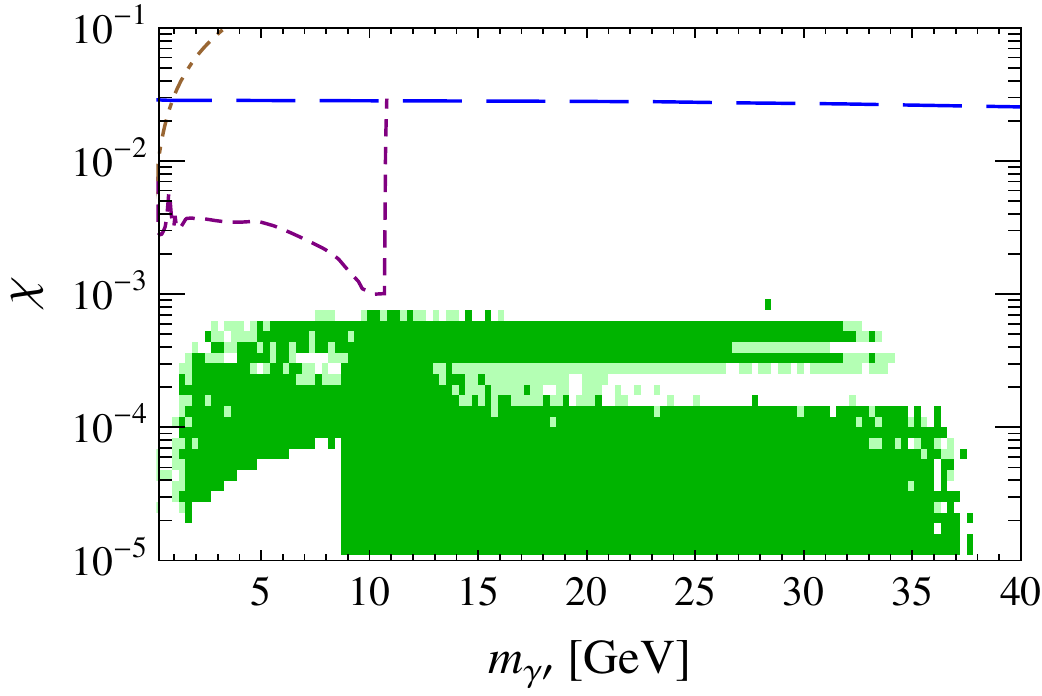}
\includegraphics[width=0.49\textwidth]{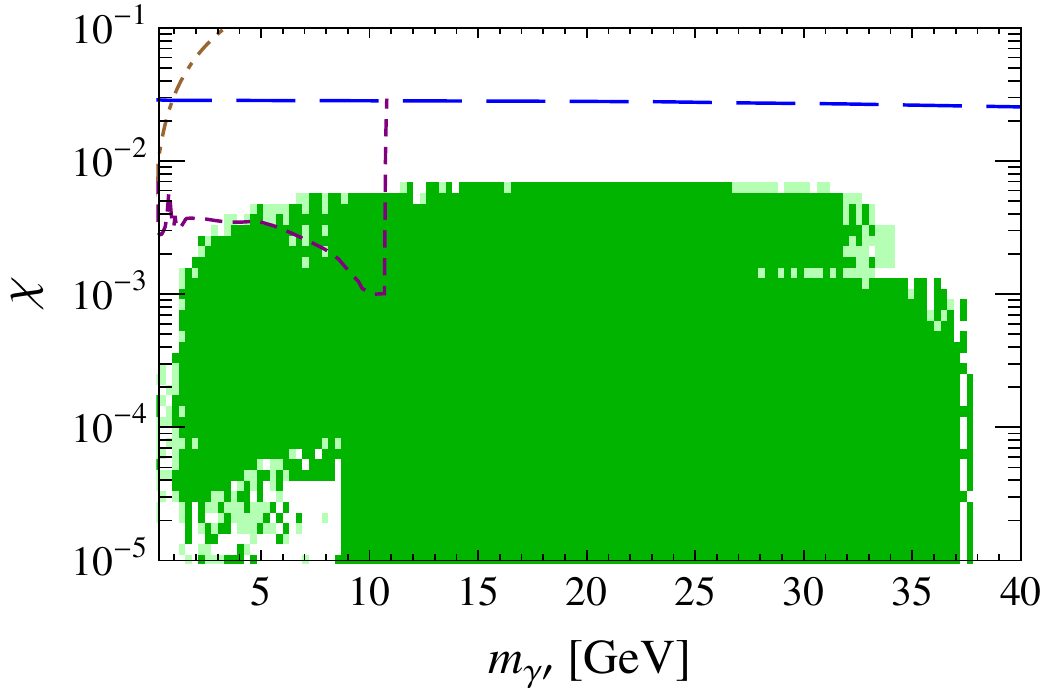}
%\marginnote{\hspace{-1cm}size increased}
\caption{Allowed space of models with radiatively induced breaking, showing hidden photon mass against the \emph{magnitude} of kinetic mixing.
Dark green areas allow for the correct dark matter relic density and
light green for subdominant dark matter. The lines represent the constraints from EWPT (long-dashed blue line),
model-dependent BaBar search (dashed dark line) and muon $g-2$ (dashed-dotted line). The Standard Halo Model (SMH) has been used,
and all DD constraints are imposed {\bf except for} the SIMPLE exclusion limit.\newline
\textit{Left:} $\kappa=1$, \textit{right:} $0.1 \leq \kappa \leq 10$.\label{RadScanNoSIMPLE}}
\end{center}
\end{figure}

\begin{figure}[p]
\begin{center}
\includegraphics[width=0.49\textwidth]{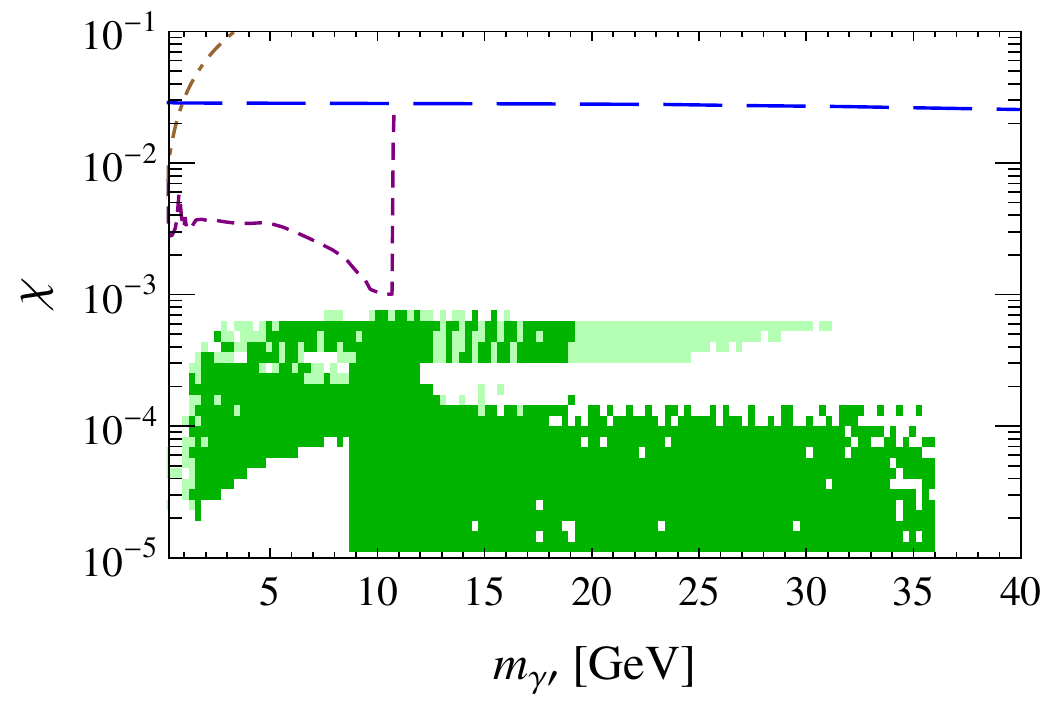}
\includegraphics[width=0.49\textwidth]{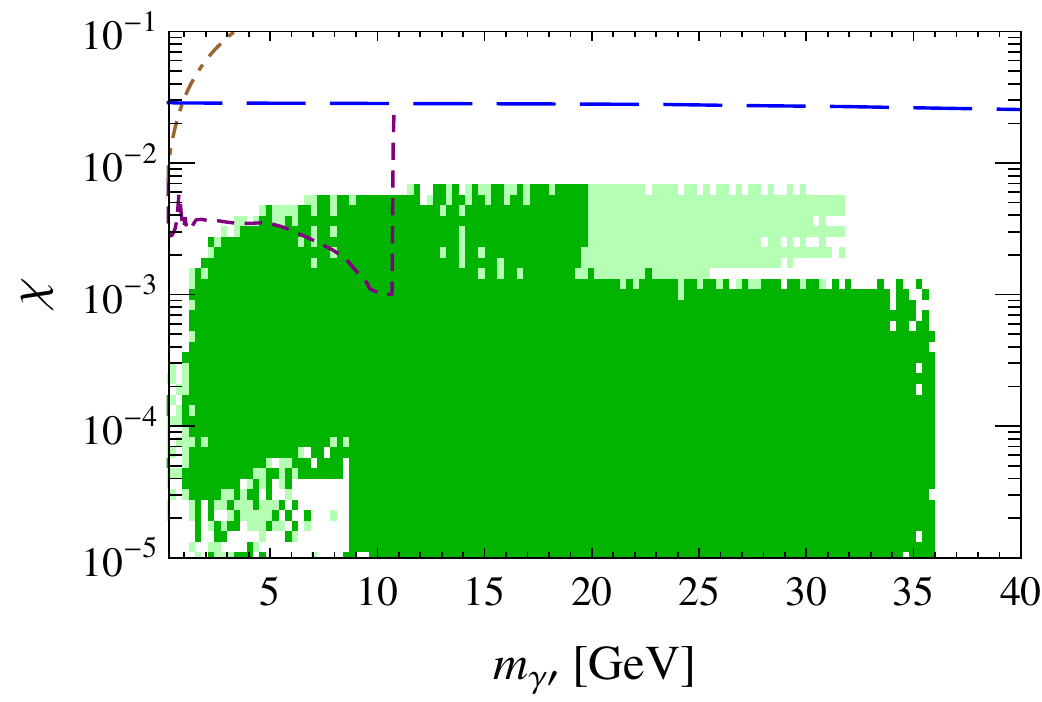}
%\marginnote{\hspace{1.7cm}size increased}
\caption{Allowed space of models with radiatively induced breaking, showing hidden photon mass against the magnitude of kinetic mixing.
Dark green areas allow for the correct dark matter relic density and
light green for subdominant dark matter. The lines represent the constraints from EWPT (long-dashed blue line),
model-dependent BaBar search (dashed dark line) and muon $g-2$ (dashed-dotted line). The Standard Halo Model (SMH) has been used,
and all DD constraints are imposed {\bf including} the SIMPLE exclusion limit.\newline
\textit{Left:} $\kappa=1$, \textit{right:} $0.1 \leq \kappa \leq 10$.\label{RadScanWithSIMPLE}}
\end{center}
\end{figure}

We then input the results of the scan into micrOMEGAs to obtain
the corresponding relic abundance and scattering cross
sections. Results for kinetic mixing against hidden photon mass
are shown in figures \ref{RadScanNoSIMPLE} and
\ref{RadScanWithSIMPLE} for $\kappa = 1$ and for a scan over
$\kappa$ in the range $0.1$ to $10$. Depending on whether the
relic abundance corresponds to the total DM abundance, the
points are shown in dark green or in light green if it is
subdominant. Clearly, allowing for variation in $\kappa$ has a
large effect on the allowed space of models, in stark contrast
to the toy model in section \ref{SEC:TOY} where the parameter
space could be filled without varying $\kappa$.


\begin{figure}[tb!]
\begin{center}
\includegraphics[width=0.49\textwidth]{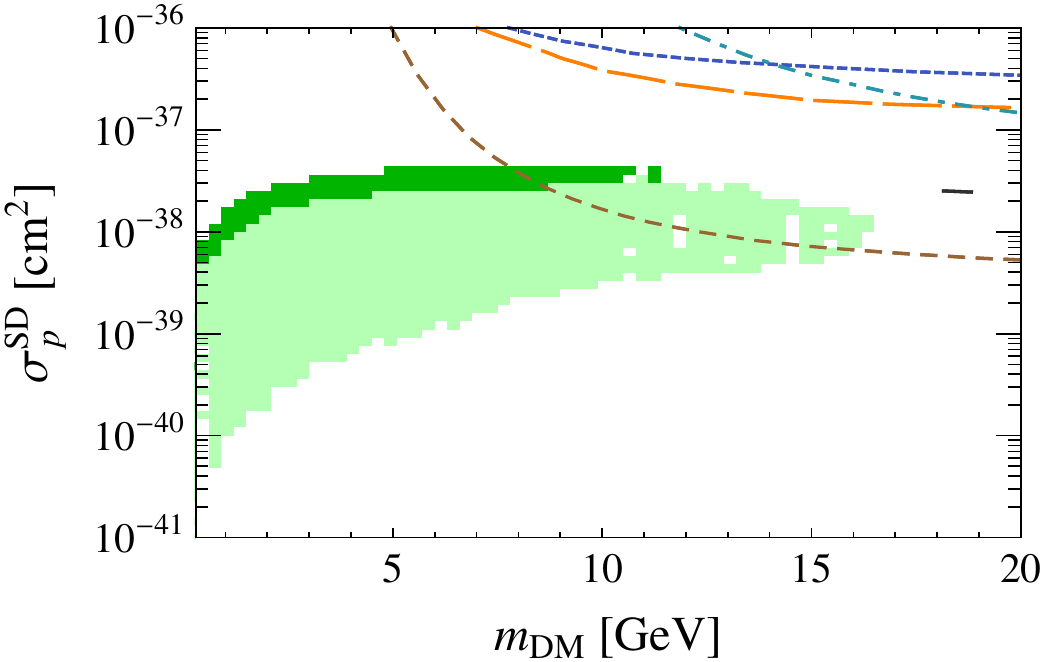}
\end{center}
\begin{center}
\includegraphics[width=0.49\textwidth]{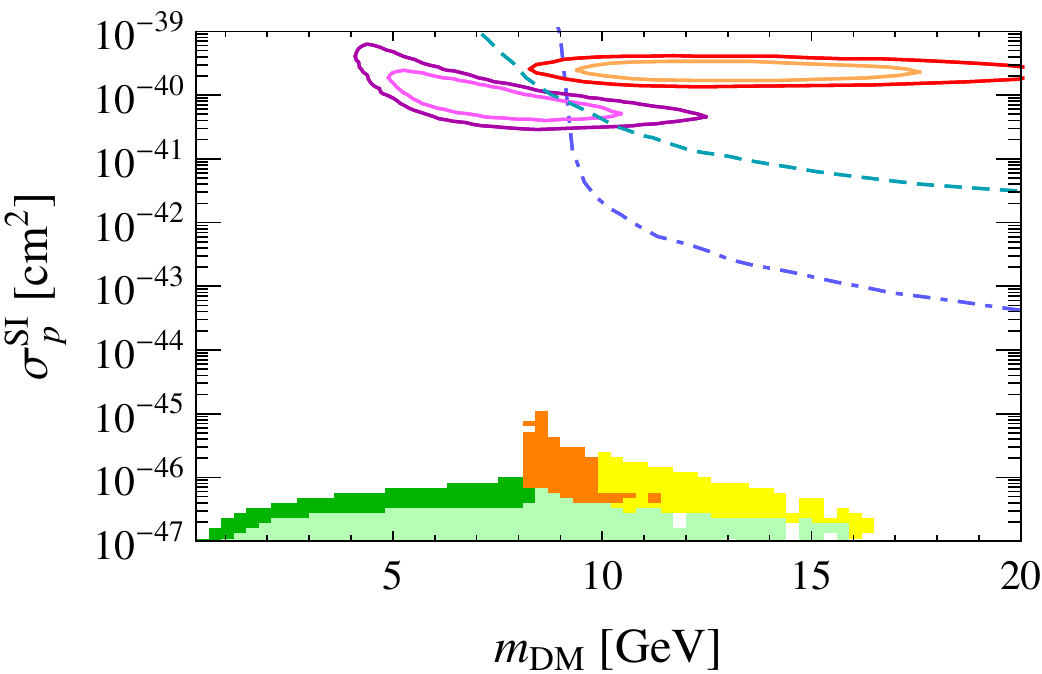}
\includegraphics[width=0.49\textwidth]{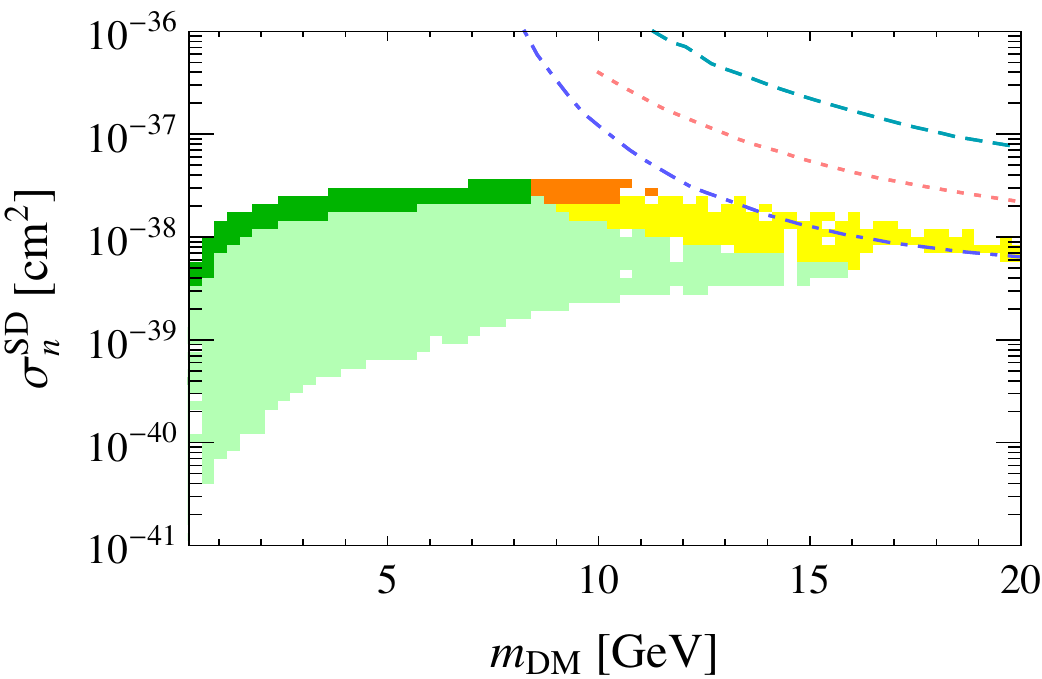}
%\marginnote{\hspace{-1cm}size increased}%
\caption{Direct detection cross sections for radiatively induced
breaking where the DM candidate is a Majorana fermion, using
SMH, scanning over $0.1 \leq \kappa \leq 10$. Dark green areas allow for the correct dark matter relic density and light green for subdominant dark matter.\textit{Top:}
spin-dependent scattering cross section ($\sigma_p^{\rm SD}$)
on protons with experimental exclusion contours: SIMPLE is shown as the
lowest-lying, short-dashed brown curve, above it is the PICASSO long-dashed orange
line, there above the COUPP2011 dashed-dotted turquoise limit, then the COUPP2007
dotted blue line and at the right of the plot starts the dashed black Super-K limit. \textit{Left bottom:} spin-independent scattering
cross section on protons ($\sigma_p^{\rm SI}$) together with
signal contours from CoGeNT (purple lines) and DAMA (red lines) as well as
exclusion limits from CDMS (dashed turquoise line) and XENON100 (dashed-dotted blue line).
\textit{Right bottom:} spin-dependent scattering
cross section on neutrons ($\sigma_n^{\rm SD}$) together with
limits from XENON10 (dashed-dotted blue line), Zeplin (dotted pink line) and CDMS (dashed turquoise line). In both plots on the bottom, points in yellow and orange lie above the SIMPLE limit while giving a subdominant and total DM abundance, respectively.
\label{RadScat}}
\end{center}
\end{figure}

The spin-dependent and spin-independent direct detection cross
sections are shown in figure~\ref{RadScat}, where $\kappa$ has
been scanned over  and we  again use the rescaling procedure
for subdominant dark matter. The dark matter particle
considered in this subsection $\tilde{o}_1$ is a Majorana
fermion and therefore has greatly suppressed spin-independent
scattering on nuclei (so there is little chance of explaining
the DAMA or CoGeNT signals via spin-independent scattering with
such a model)\footnote{Here, we do not study different halo
profiles as they only effect the potential signals in direct
detection experiments for which the required cross sections are
several orders of magnitude above those obtained in the models
of subsection~\ref{SEC:RADBREAK}. Therefore, the results
presented for the SMH do not differ for other halo models.};
however, the obtained spin-dependent cross sections are quite
large, and some are even already excluded by current experiments
(the experiments were mentioned in section
\ref{SEC:CONSTRAINTS:DM}). The most stringent constraint arises
from the SIMPLE experiment for SD scattering on protons which cuts out many parameter points
for dark matter masses above $\sim 6$ GeV; the effect of this
is illustrated by showing the parameter scan before SIMPLE is
included in figure~\ref{RadScanNoSIMPLE} and afterwards in
figure~\ref{RadScanWithSIMPLE}. Since the spin-dependent and spin-independent cross sections are related, the SIMPLE limit removes a large portion of the parameter space with larger values of \emph{spin-independent} scattering direct-detection cross sections. This is illustrated in figure~\ref{RadScat}, which presents the different scattering cross sections. The top plot contains all the spin-dependent cross sections on protons with the experimental bounds. At the bottom, the corresponding spin-independent cross sections are shown on the left and the spin-dependent ones on neutrons on the right. In those two plots, yellow and orange points indicate models that lie in the top plot above the SIMPLE exclusion contour for spin-dependent scattering on protons (points in yellow have a subdominant and points in orange the total DM abundance). As can be seen from the right bottom plot, the SIMPLE limit for spin-dependent scattering on protons is more constraining than limits from scattering on neutrons where XENON10 can exclude only few models.

As mentioned above, the fact that the hidden sector dark matter
candidate is a Majorana fermion leads to extremely small
spin-independent scattering cross sections. They do, however,
obtain a contribution from the Higgs portal term, which in
supersymmetric theories is always present. We describe this in
detail in appendix \ref{APP:HIGGS}, where we also derive a
simple approximation for the contribution of the Higgs portal
term which agrees well with the results seen in figure
\ref{RadScat}:
\begin{align}
\sigma_N^{\rm SI,\, Portal} \sim& 10^{-45} \mathrm{cm}^2 \times \left(\frac{m_{\tilde{o}_1} }{m_N +  m_{\tilde{o}_1}}\right)^2 \left( \frac{\chi s_W}{0.001} \right)^2  \left(\frac{\mathrm{GeV}}{m_{\gamma'}} \right)^2,
\end{align}
There is also a somewhat smaller and more spectrum-dependent
contribution from squark exchange. For the Majorana fermion DM
of this section, the spin-independent nuclear cross sections
are very similar for scattering on protons and on neutrons;
hence, we have written $\sigma_N^{\rm SI,\, Portal} $ with ``N''
to denote Nucleons; in the plots (figure \ref{RadScat}) the cross sections on protons %\marginnote{fig7} %
($\sigma_p^{\rm SI,\, Portal}$) are shown, which also allows
direct comparison with the next subsection.

Our results in this subsection are largely independent of the
halo model applied to the spin-independent scattering limits,
as the corresponding cross sections are much below the
experimental reach.

\subsubsection{Example model}

To better understand the types of models that we find, since
the plots can only show two-dimensional parameter spaces, here
we give an example of one of the models that satisfies all
experimental constraints and provides the entire dark matter
density.  We take $\kappa$ to be unity and the soft masses
$m_{H_\pm}$ approximately $100$ GeV at the high-energy scale.
We then run the parameters down and adjusted at the high scale
to find appropriate values at low energies; thus, $m_S$ is
somewhat larger and drives the soft hidden Higgs masses to
become tachyonic. The parameters at low ($10$ GeV) and high
($10^{16}$ GeV) energy scales are given in table
\ref{TAB:GRAVMODEL} along with the spectrum at low energies
after hidden gauge symmetry breaking. The dark matter candidate
is then the Majorana fermion $\tilde{o}_1$, having a mass of
$5.2$ GeV and yielding a density of $\Omega_{\tilde{o}_1} h^2 =
0.112$. The spin-independent nuclear direct detection
cross section is $\sigma_p^{\rm SI}= 3.6 \times 10^{-47}
\cm^2$, the spin-dependent cross section being $\sigma_p^{\rm
SD}=2.5\times 10^{-38} \cm^2$. The mass of the hidden photon
and hidden Higgs is  $11.6$ GeV, and they have widths of  $6.7
\times 10^{-8}$ GeV and $4.8 \times 10^{-8}$ GeV, respectively,
the latter decaying mostly to charm and $b$ quarks.

\begin{table}[htb!]
\begin{center}
$\begin{array}{||c|c||}\hline\hline\multicolumn{2}{||c||}{\mathrm{High\ scale\ parameters}}  \\\hline
\kappa & -1.0\\
\chi & -0.0008\\
\alpha_h  &   0.0031\\
\alpha_S &    0.011\\
M_\lambda &    21.4\  \mathrm{GeV}\\
m_{H_+}^2 &    (101)^2\ \mathrm{GeV}^2\\
m_{H_-}^2 &   (101)^2\ \mathrm{GeV}^2\\
m_S^2 &   (418)^2\  \mathrm{GeV}^2 \\
A_S &    -0.2\  \mathrm{GeV}\\\hline\hline
\end{array}$
$\begin{array}{||c|c||}\hline\hline\multicolumn{2}{||c||}{\mathrm{Low\ scale\ parameters}} \\\hline
\kappa  &   -1.0 \\
\chi &    -0.0005\\
\alpha_h  &   0.003\\
\alpha_S &    0.010\\
M_\lambda &    20.7\ \mathrm{GeV}\\
m_{H_+}^2 &    -66.8\  \mathrm{GeV}^2\\
m_{H_-}^2 &   -68.9\  \mathrm{GeV}^2\\
m_S^2 &    (406)^2\  \mathrm{GeV}^2\\
A_S &    -1.5\  \mathrm{GeV}\\\hline\hline
\end{array}$
$\begin{array}{||c|c||}\hline \hline \mathrm{Particle} & \mathrm{Mass\ (GeV)} \\\hline
\tilde{o}_7  &   14.0\\
\tilde{o}_1  &   5.2\\
\tilde{o}_2  &   25.9\\
\gamma^{\prime} &    11.6\\
H_+ &    11.6\\
H_-,\ S & 7.7, 406\\\hline
\end{array}$
\caption{Hidden sector parameters and particle masses for an
example gravity mediated model, yielding the entire dark matter
density $\Omega_{\tilde{o}_1} h^2 = 0.112$. The direct
detection nucleon cross sections are $\sigma_p^{\rm SI}=
3.6\times 10^{-47} \cm^2$ and $\sigma_p^{\rm SD}=  2.5\times
10^{-38} \cm^2$, well outside the reach of current
experiments.} \label{TAB:GRAVMODEL}\end{center}
\end{table}

For the SPS1b data point~\cite{Allanach:2002nj}, the full
neutralino mass matrix  in the basis\\ $(B_0, W_0, h_u^0,
h_d^0, \tilde{\lambda},\tilde{h}_+)$ is (in GeV)
\begin{equation}
\mathcal{M}_{\rm neutralino}= \left(\begin{array}{cccccc}
 166 & 0 & -2.73 & 43.8 & -0.01 & -0.01 \\
0 & 310 & 2.73 & -79.9 & 0 & 0 \\
-2.73 & 2.73 & 0.00 & -511 & 0 & 0\\
43.8 & -79.9 & -511 & 0 & 0 & 0 \\
-0.01 & 0 & 0 & 0 & 20.7 & 11.6 \\
-0.01 & 0 & 0 & 0 & 11.6 & 0 \end{array}\right)
\end{equation}
with eigenmasses $5.2, 25.9, 164, 298, 516$ and $530$ GeV. For the same data point, we can compute the mixing between the original Higgs eigenstates and the mass eigenstates to be
\begin{equation}
\vect3[H_+,h,H] = \arrb3[1.0,-3.6\times 10^{-5},1.2\times 10^{-7}][3.6\times 10^{-5},1.0, 0][-1.2\times 10^{-7},0 ,1.0] \vect3[h_1, h_2, h_3].
\end{equation}

\subsection{Visible sector induced breaking}

\subsubsection{Parameter scans}

\begin{figure}[htb!]
\begin{center}
\includegraphics[width=0.49\textwidth]{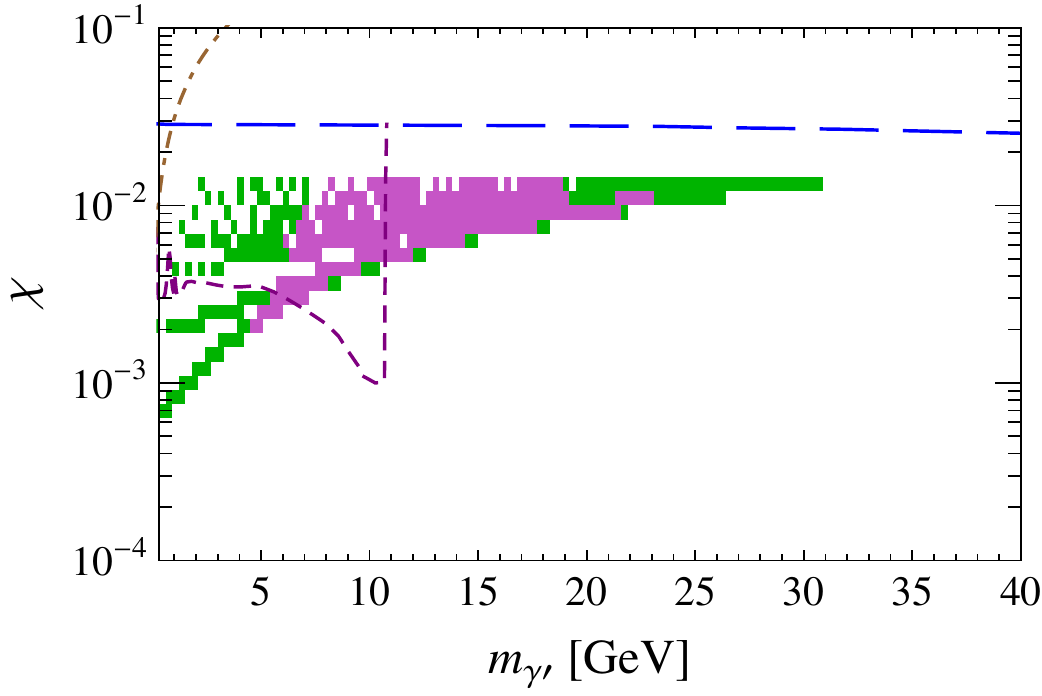}
\includegraphics[width=0.49\textwidth]{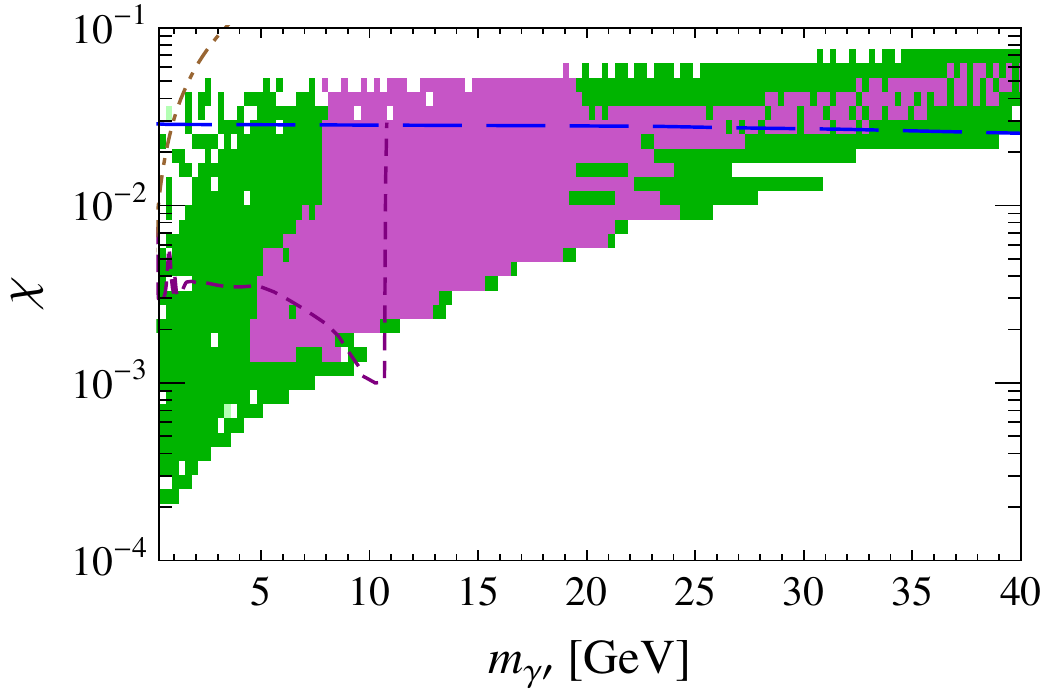}
%\marginnote{\hspace{1.7cm}size increased}%
\caption{Allowed space of models with visible-sector induced breaking as a function of the hidden photon
mass and the magnitude of kinetic mixing,
showing dark green areas where the correct dark matter relic abundance can be found
and light purple where the CoGeNT signal can be explained with a subdominant dark matter candidate.
The lines represent the constraints from EWPT (long-dashed blue line),
model-dependent BaBar search (dashed dark line) and muon $g-2$ (dashed-dotted line).
The Standard Halo Model (SMH) has been used, and all DD constraints are imposed including the SIMPLE exclusion limit.
\textit{Left:} $\kappa=1$, \textit{right:} $0.1 \leq \kappa \leq 10$. \label{VisibleShowKappa}}
\end{center}
\end{figure}

Here, we implement a scan for visible sector induced breaking,
by scanning over parameters  at the low-energy scale. As in the
previous subsection, we insist on perturbativity for
$\lambda_S$ and $g_h$ and take a maximum value of
$m_{\gamma'}$ of $40$ GeV. However, the soft supersymmetry
breaking masses are chosen to be small, as may be induced in
gauge mediation or sequestering of the hidden sector.
Phenomenologically, then, the results of this subsection can be
considered to be a detailed examination of the model of
\cite{Morrissey:2009ur}, but with a large gravitino mass and
kinetic mixing respecting the relation (\ref{ChiRelation}).

\begin{figure}[htb!]
\begin{center}
%\includegraphics[width=7cm]{PlotsGaugeMed/ScatPlot_chivsmgamma_GaugeMed_NFW_resc_final_allkappa.pdf}
%\end{center}
%\begin{center}
\includegraphics[width=0.49\textwidth]{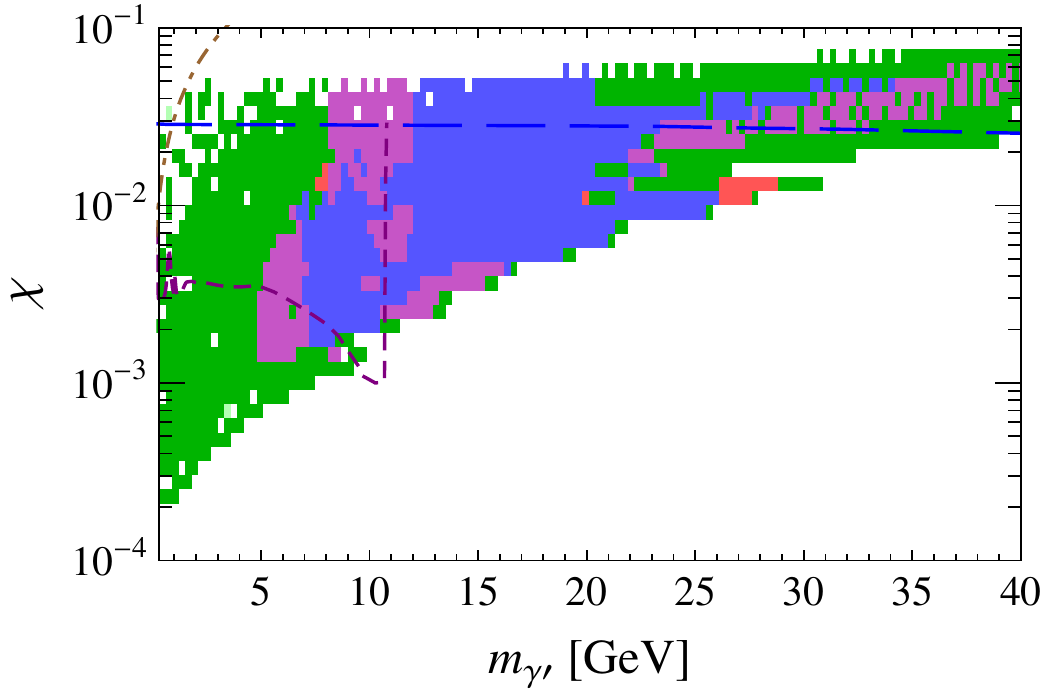}
\includegraphics[width=0.49\textwidth]{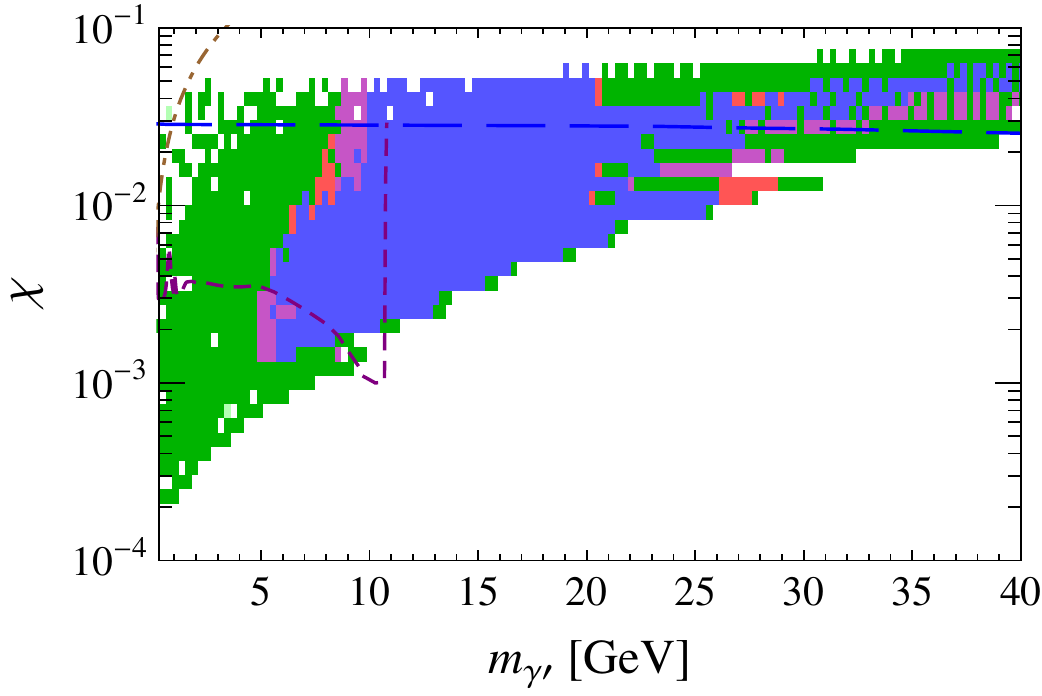}
%\marginnote{\hspace{-1.5cm}plot NFW removed}
\caption{Allowed space of models with visible-sector induced breaking in function of the hidden photon
mass and the magnitude of kinetic mixing, scanned over $0.1 \leq \kappa \leq 10$,
showing different halo models: \textit{left:} Isothermal halo model;
\textit{right:} Einasto halo model. Here, the red region shows the space explaining the DAMA signal,
the purple region explains the CoGeNT signal, and the blue one explains both DAMA and CoGeNT,
all signal regions having a subdominant dark matter density. All DD constraints including SIMPLE are imposed.
The lines represent the constraints from EWPT (long-dashed blue line),
model-dependent BaBar search (dashed dark line) and muon $g-2$ (dashed-dotted line).}
\label{VisibleShowHalos}
\end{center}
\end{figure}

As mentioned in section \ref{SEC:DMCANDIDATES}, we can have
either of two dark matter candidates, depending on the
particular low-energy parameters: either the Majorana fermion
$\tilde{o}_1$ or a Dirac fermion $\tilde{o}_7$. For both cases,
we again use micrOMEGAs to compute the relic abundance and the
scattering cross sections. The space of models in the kinetic
mixing--hidden photon mass plane (all points shown are in
agreement with all direct detection exclusions, including
SIMPLE) is shown in figures~\ref{VisibleShowKappa}
and~\ref{VisibleShowHalos}; the colour code is identical to the
scatter plots for the toy model in figure~\ref{Fig:ToyModScatt},
and the different experimental constraints are explained in
sections~\ref{SEC:CONSTRAINTS:HP} and~\ref{SEC:CONSTRAINTS:DM}.
Figure~\ref{VisibleShowKappa} demonstrates the expansion in the
parameter space by allowing a variation in $\kappa$; both
$\kappa=1$ (left plot) and $0.1 \le \kappa \le 10$ (right plot)
are shown for the Standard Halo Model (SMH). The effect of
changing the halo model is illustrated in
figure~\ref{VisibleShowHalos}. Depending on the halo model we
find subdominant DM explanations for DAMA and CoGeNT
separately, as well as for both simultaneously, which are
represented as light red, purple and blue regions, respectively.
The two experiments can only be explained simultaneously for
certain halo models other than the SMH.

The resulting parameter points in figures
\ref{VisibleShowKappa}   and \ref{VisibleShowHalos} show a very
similar behavior to the toy model (see figure
\ref{Fig:ToyModScatt}) as here the dark matter candidate can
also be a Dirac fermion. The main difference is that the models
here never permit annihilation via the $t$-channel diagram --
since the dark matter particle can never be heavier than the
hidden gauge boson. Therefore, the lower part of the plots  is,
in contrast to the toy model, empty as there it was filled by
dark green points finding the correct relic abundance  lying
either just above the threshold for $t$-channel annihilation or
on the $s$-channel resonance (as we scan over the dark matter
mass, these resonances  move through the plot through different
values of $m_{\gamma'}$, and the whole range is covered). The
coarser grid and small holes in the current scatter plots
compared to the toy model arise from the fact that the
parameter space can not be scanned as continuously as for the
toy model.

\begin{figure}[p]
\begin{center}
\includegraphics[width=0.49\textwidth]{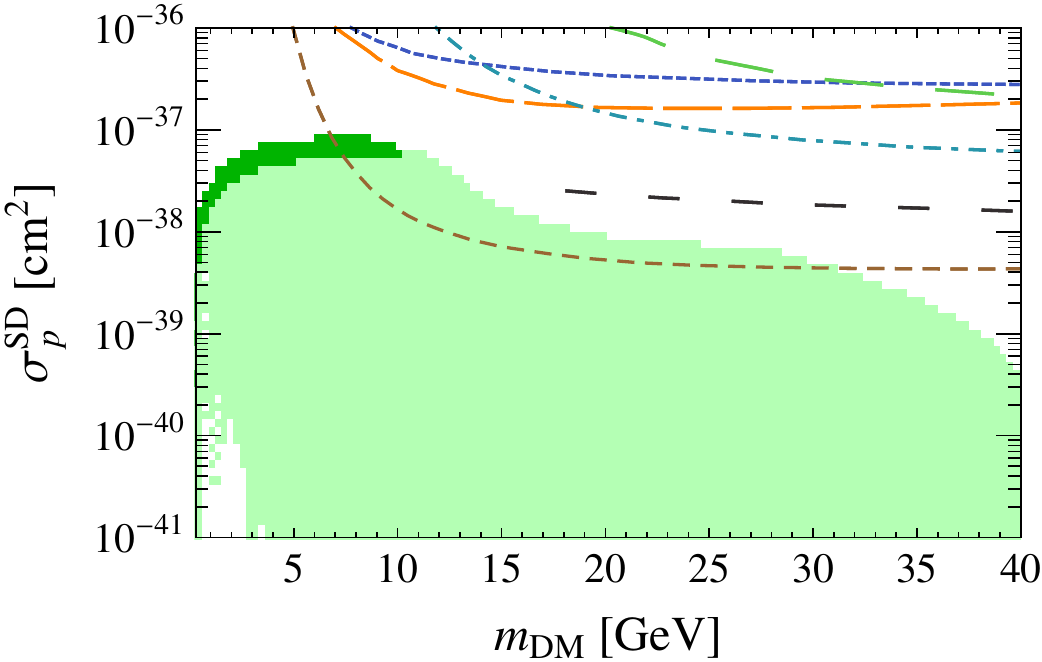}
\includegraphics[width=0.49\textwidth]{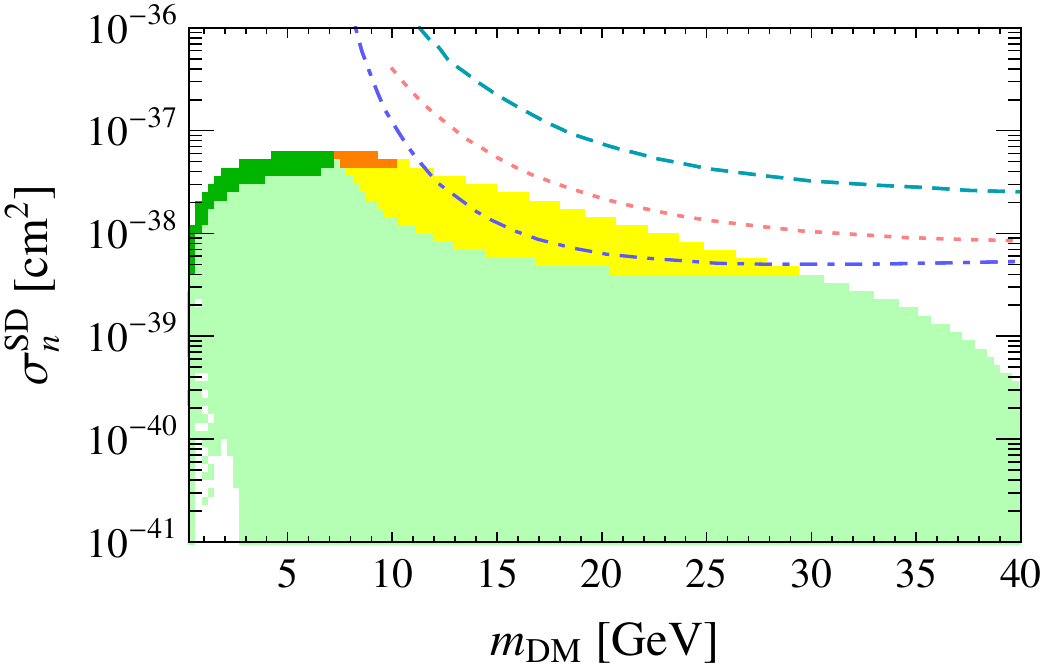}
%\marginnote{\hspace{1.7cm}size increased}
\caption{Spin-dependent scattering cross sections on protons (\textit{left}) and on neutrons (\textit{right}) for visible-sector induced breaking together
with the exclusion contours from the corresponding direct detection experiments
(the lowest lying, short-dashed brown line in the left plot is the SIMPLE limit, the other lines are as explained in figure~\ref{RadScat},
and in addition the long-dashed green line is the KIMS limit).
Here, only the Majorana dark matter candidate $\tilde{o}_1$ is shown
(the cross sections for the Dirac fermion $\tilde{o}_7$ are too small to appear). In the right plot, points in yellow and orange lie above the SIMPLE limit while giving a subdominant and total DM abundance respectively.
The SMH has been used, and $0.1 \leq \kappa \leq 10$.\label{VisibleSDMajorana}}
\end{center}
\end{figure}

\begin{figure}[p]
\begin{center}
\includegraphics[width=0.49\textwidth]{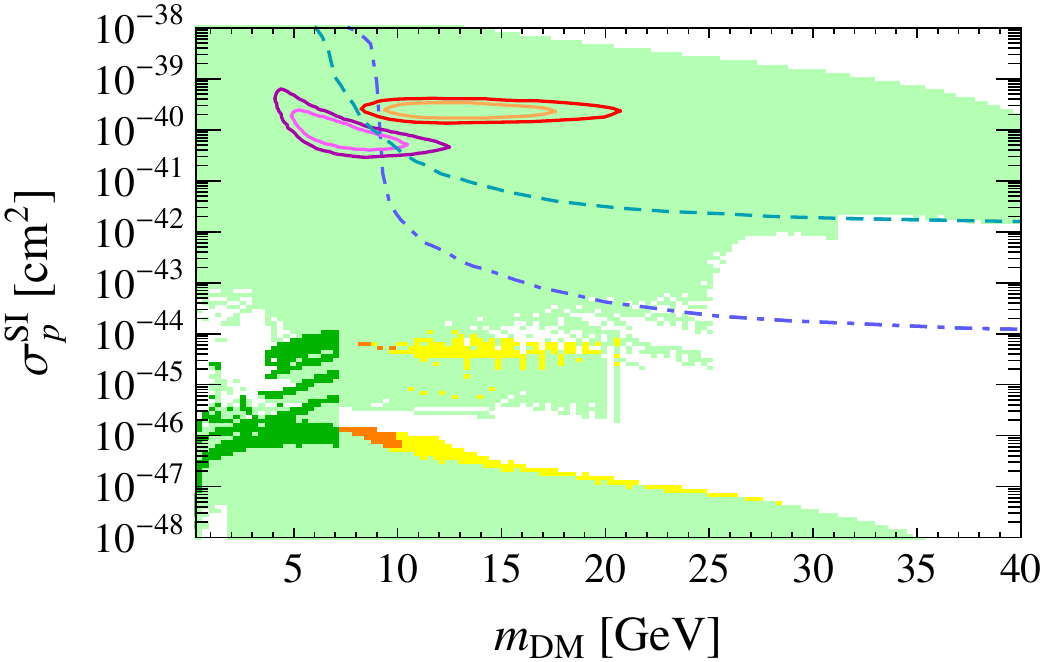}
\includegraphics[width=0.49\textwidth]{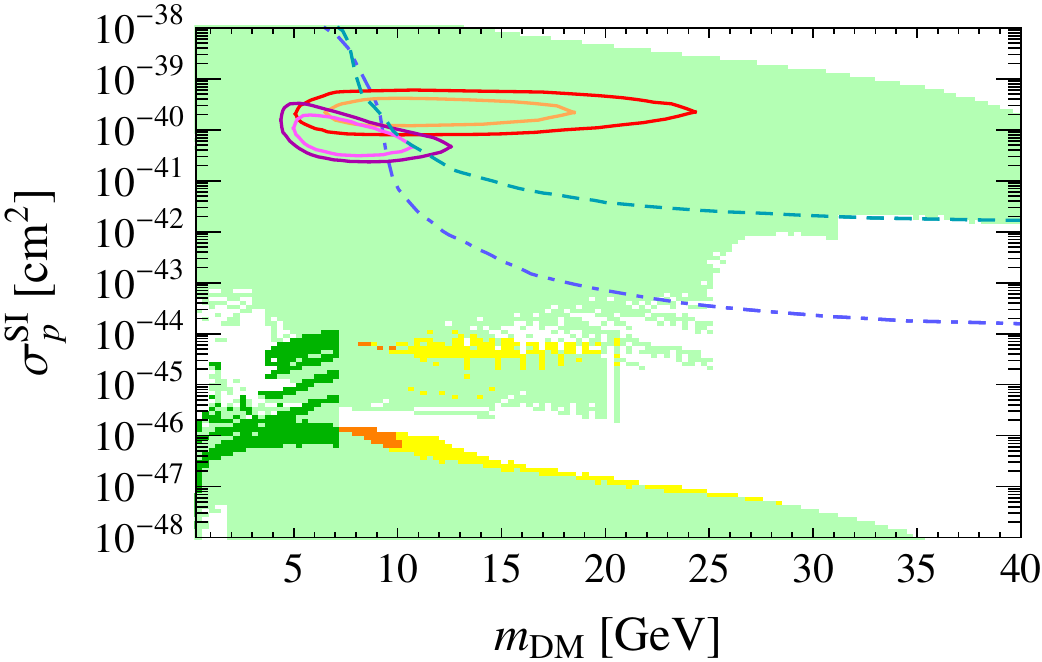}
%\marginnote{\hspace{1.7cm}size increased}
\caption{Spin-independent scattering cross sections for visible sector induced breaking using
(\emph{left}) the SMH and (\emph{right}) the Einasto profile as given in \cite{Arina:2011si} and $0.1 \leq \kappa \leq 10$.
The signal contours from CoGeNT (purple lines) and DAMA (red lines) are shown, which overlap for the right-hand plot;
the exclusion limits from CDMS and XENON100 are shown as dashed turquoise and dashed-dotted blue curves, respectively.
The plot splits into two disjoint green areas: in the upper, the Dirac fermion $\tilde{o}_7$
is the dark matter candidate, while in the lower one, it is the Majorana fermion $\tilde{o}_1$. In both plots, points in yellow and orange lie above the SIMPLE limit while giving a subdominant and total DM abundance, respectively.
\label{VisibleSI}}
\end{center}
\end{figure}

The spin-dependent and spin-independent scattering cross
sections for the Standard Halo Model are shown in
figure~\ref{VisibleSDMajorana} and~\ref{VisibleSI},
respectively. In both figures, the effect of including the SIMPLE limit is shown: points in yellow and orange indicate models whose SD scattering cross section on protons is excluded by SIMPLE while giving a subdominant and total DM abundance, respectively. The
spin-dependent scattering cross sections in
figure~\ref{VisibleSDMajorana} are only appreciable when the
Majorana fermion $\tilde{o}_1$ is the dark matter candidate. In
figure~\ref{VisibleSI} for SI scattering, there are two disjoint
regions corresponding to whether the dark matter candidate is
the Majorana fermion $\tilde{o}_1$ (lower region) or the Dirac
fermion $\tilde{o}_7$ (upper region). As in the
radiatively-induced breaking case, the Majorana fermion has a
small spin-independent cross section of $10^{-47}$ to
$10^{-45}\cm^2$. In contrast to this, the spin-independent
scattering cross section of the Dirac fermion $\tilde{o}_7$ is
in the range of current direct detection experiments and may
explain the signals in CoGeNT and DAMA via a subdominant dark
matter component. Hence, we present the effect of changing the
halo model on the CoGeNT and DAMA regions in
figure~\ref{VisibleSI}, showing that simultaneous
explanations of both signals are possible and justify the blue
regions in figure~\ref{VisibleShowHalos}.


The Dirac fermion $\tilde{o}_7$ has almost no spin-dependent
scattering on nuclei, so the SIMPLE exclusion limit only
affects the lower regions in the plots for spin-independent
scattering (figure~\ref{VisibleSI}) which correspond to the
Majorana fermion, while the parameter regions that are
interesting for spin-independent scattering experiments and can
explain the DAMA and CoGeNT signals remain untouched. The SI
scattering cross sections plotted are those for
scattering on protons. As
described above, the Dirac fermion's interaction through
kinetic mixing couples almost exclusively to the charge of the
nucleon, so its SI scattering on neutrons essentially vanishes.
For the Majorana fermion $\tilde{o}_1$ on the other hand,
spin-independent scattering on protons and neutrons is of
roughly equal magnitude as it proceeds via the Higgs portal and
squark exchange (as in the previous subsection), but the plots (figure \ref{VisibleSI})   %\marginnote{~fig 11} %
also show only the cross section on protons. As shown in figure \ref{VisibleSDMajorana}, the spin-dependent
scattering cross sections of the Majorana fermion DM candidate are also almost the same for protons and neutrons. However, the SIMPLE limit on the former (left plot) is slightly more constraining than the XENON 10 limit on the latter (right plot). In the case of the
Dirac fermion dark matter, the spin-dependent scattering
essentially vanishes both for protons and neutrons.

\subsubsection{Example model}

An example of a model that can explain the DAMA and CoGeNT
signals (when we use the Einasto profile) is given in table
\ref{TAB:SEQMODEL} where the spectrum is given. We take, at the
low-energy scale, $\kappa = -10, \alpha_h (=\frac{g_h^2}{4\pi})
= 0.040, \alpha_S (=\frac{\lambda_S^2}{4\pi}) =0.027$ (giving
$\chi = -0.016$) and the soft masses given by sequestered
values $M_\lambda = m_{H_+}^2 = m_{H_+}^2 =m_S^2 = 1$ GeV, with
the hidden A-term $A_S =0$. The dark matter candidate is
necessarily then $\tilde{o}_7$, having a mass of $6.4$ GeV and
yielding a density of $\Omega_{\tilde{o}_7} h^2 = 0.0021$. The
cross section is almost entirely on protons, which, when
rescaled to the dark matter density, yields an effective
scattering cross section of $\sigma_p^{\rm SI}= 1.0 \times
10^{-40} \cm^2$. The hidden photon mass is $11$ GeV, with width
$3.7\times 10^{-5}$ GeV, decaying mostly into light leptons and
quarks. The hidden Higgs width is $2\times 10^{-10}$ GeV,
mostly decaying to charm and $\tau$s.

\begin{table}[htb!]
\begin{center}
$\begin{array}{||c|c||}\hline\hline\multicolumn{2}{||c||}{\mathrm{Low\ scale\ parameters}} \\\hline
\kappa  &   -10\\
\chi &    -0.016\\
\alpha_h  &   0.040\\
\alpha_S &    0.027\\
M_\lambda &    1.0\ (\mathrm{GeV})^2\\
m_{H_+}^2 &    1.0\ (\mathrm{GeV})^2\\
m_{H_-}^2 &   1.0\ (\mathrm{GeV})^2\\
m_S^2 &    1.0\ (\mathrm{GeV})^2\\
A_S &    0.0\\ \hline\hline
\end{array}$
$\begin{array}{||c|c||}\hline \hline \mathrm{Particle} & \mathrm{Mass\ (GeV)} \\\hline
\tilde{o}_7  &   6.4\\
\tilde{o}_1  &   10.5\\
\tilde{o}_2  &   11.5\\
\gamma^{\prime} &    11.0\\
H_+ &    11.0\\
H_-,\ S & 6.4, 6.5\\\hline
\end{array}$
\caption{Hidden sector parameters and particle masses for an
example sequestered model, yielding a dark matter density
$\Omega_{\tilde{o}_7} h^2 = 1.7 \times 10^{-3}$ and rescaled
direct detection cross section for scattering on protons
$\sigma_p^{\rm SI}= 1.0 \times 10^{-40} \cm^2$.}
\label{TAB:SEQMODEL}\end{center}
\end{table}

For the SPS1b data point \cite{Allanach:2002nj}, the full
neutralino mass matrix  in the basis\\ $(B_0, W_0, h_u^0,
h_d^0, \tilde{\lambda},\tilde{h}_+)$ is
\begin{equation}
\mathcal{M}_{\rm neutralino}= \left(\begin{array}{cccccc}
  166 & 0.00 & -2.73 & 43.8 & -0.02 & -0.18 \\
0.00 & 309 & 2.73 & -79.9 & 0.00 & 0.00 \\
-2.73 & 2.73 & 0.00 & -511 & 0.00 & 0.00 \\
43.8 & -79.90 & -511 & 0.00 & 0.00 & 0.00 \\
-0.02 & 0.00 & 0.00 & 0.00 & 1.00 & 11.0 \\
-0.18 & 0.00 & 0.00 & 0.00 & 11.0 & 0.00
\end{array}\right)
\end{equation}
with eigenmasses $10.5, 11.5, 164, 298, 516$ and $530$ GeV. For
the same data point, we can compute the mixing between the
original Higgs eigenstates and the mass eigenstates to be
\begin{equation}
\vect3[H_+,h,H] = \arrb3[1.0,-1.2\times 10^{-3},3.9\times 10^{-6}][1.2\times 10^{-4} ,1.0, 0][-3.9\times 10^{-6},0 ,1.0] \vect3[h_1, h_2, h_3].
\end{equation}

\section{Conclusions and outlook}
\label{SEC:CONCLUSIONS}

We have presented what we believe to be the first detailed
examination of the dark matter relic abundance and direct
detection cross sections of a complete string-inspired supersymmetric dark
force model, emphasising the natural supersymmetric
relationship between kinetic mixing and the hidden gauge
coupling. In particular, we have included running from high-energy gravity-mediated boundary conditions and shown that
interesting and viable models exist, in contrast to prior
expectations. We have also examined the effect of neutralino
mixing and the Higgs portal term, showing that the latter can
contribute a small spin-independent cross section for Majorana
fermion dark matter candidates. We  examined the model in the
cases of both radiative and visible sector induced hidden gauge
symmetry breaking and demonstrated the stark phenomenological
contrasts between the two.

While the model can be used to explain the current dark matter
signals observed by DAMA, CoGeNT and CRESST, this is not
plausible in the case of radiative-induced breaking relevant
for gravity mediation, where our motivation was to show that
simple dark sectors are not excluded -- the hidden \U1 may
instead be detected in fixed target experiments, particularly
if the hidden photon cannot decay to hidden matter (as in the
reasonably generic case when the dark matter particle has mass
near that of the hidden photon). However, this is certainly
plausible if the model were to be extended, for example, by
allowing a supersymmetric mass for the singlet.

We hope that this work has paved the way for more detailed
analysis of other supersymmetric dark sectors. In addition,
there are several further possible avenues of work within the
current model, for example, by including the full loop
corrections to the effective potential, and constraints from
indirect dark matter searches (which is a work in
progress~\cite{inPrep}, although we believe them to be less
stringent than the direct searches). It would also be
interesting to compare the signal from CRESST with those from
DAMA and CoGeNT in the context of these models.

\section*{Acknowledgments}

MDG was supported by SFB grant 676 and by the ERC advanced grant 226371. He would like to thank
Genevi\`eve Belanger for helpful discussions and information
about micrOMEGAs and James Wells for discussions about BBN. SA  thanks Chiara Arina for helpful
discussions on direct detection. All of the authors would like
to thank Yann Mambrini for informative discussions about his
related work, Andreas Goudelis for useful conversations about
indirect detection and J\"org J\"ackel and Sa\'ul Ramos-S\'anchez
for interesting conversations.

\appendix

\section{Renormalisation group equations}
\label{APP:RGEs}

Here, we present the two-loop renormalisation group equations
for the hidden-sector parameters $\alpha_S \equiv
\lambda_S^2/4\pi, \alpha_h = g^2_h/4\pi, M_\lambda, m_+^2,
m_-^2, m_S^2, A_S$. We
define $t \equiv \log \mu$:
\begin{align}
\frac{d \alpha_S}{dt} =& \frac{1}{4\pi} [ 2\alpha_S ( 3\alpha_S - 4 \alpha_h)] +\frac{1}{(4\pi)^2} [ 4\alpha_S ( -3\alpha_S^2 + 2 \alpha_S \alpha_h + 8 \alpha^2_h)] \nn\\
\frac{d \alpha_h}{dt} =& \frac{1}{4\pi} [ 4\alpha^2_h ] +\frac{1}{(4\pi)^2} [ 8\alpha^2_h ( 2\alpha_h - \alpha_S)]\nn\\
\frac{d M_\lambda}{dt} =& \frac{1}{4\pi} [ 4\alpha_h M_\lambda ] +\frac{1}{(4\pi)^2} [ 8\alpha_h M_\lambda ( 4 \alpha^2_h - \alpha_S)  + 8 \alpha_h \alpha_S A_S] \nn\\
\frac{d A_S}{dt} =& \frac{1}{4\pi} [ 6\alpha_S A_S  + 8 \alpha_h M_\lambda ] +\frac{1}{(4\pi)^2} [ 8A_S  ( \alpha_h \alpha_S - 3\alpha_S^2) - 8 \alpha_h M_\lambda (\alpha_S + 8 \alpha_h)] \nn\\
\frac{d m_S^2}{dt} =& \frac{1}{4\pi} [ 2\alpha_S (m_S^2 + m_+^2 + m_-^2 + A_S^2)] \nn\\
&+\frac{1}{(4\pi)^2} \bigg[ 8\alpha_S (\alpha_h - \alpha_S^2) (m_S^2 + m_+^2 + m_-^2 + 2A_S^2) +8 \alpha_h \alpha_S (2M_\lambda^2 - 2M_\lambda A_S - A_S^2) \bigg] \nn\\
\frac{d m_\pm^2}{dt} =& \frac{1}{4\pi} [ 2\alpha_S (m_S^2 + m_+^2 + m_-^2 + A_S^2) - 8 M_\lambda^2 \alpha_h \pm 2 \alpha_h (m_+^2 - m_-^2)] \nn\\
&+\frac{1}{(4\pi)^2} \bigg[ -8 \alpha_S^2 (m_S^2 + m_+^2 + m_-^2 + 2A_S^2) +96 \alpha^2_h  M_\lambda^2 \nn\\
&\qquad\qquad+ 8 \alpha^2_h (m_+^2 + m_-^2)  \pm 4 \alpha (2\alpha_h - \alpha_S) (m_+^2 - m_-^2) \bigg].
\end{align}

The reader may be surprised to see that the kinetic mixing or the visible sector parameters are absent, but this is perfectly correct to two-loop order. This is because the kinetic mixing is a one-loop quantity (as can be seen from the canonical Lagrangian (\ref{EQ:CANONICAL})), and we maintain the normalisation of the gauge field strengths throughout the running to be
\begin{align}
\C{L} \supset \int d^2 \theta \bigg(\frac{1}{4} B^\alpha B_\alpha + \frac{1}{4} X^\alpha
X_\alpha - \frac{\chi}{2} B^\alpha X_\alpha \bigg)
\end{align}
i.e. we do not diagonalise the gauge fields. Until we arrive at low energies, the hypercharge and hidden \U1 are massless, and so there is a continuous choice of basis for the fields. However, we have two key assumptions: first, we are assuming that the kinetic mixing is generated by a high-energy theory, above the scale at which we begin the running, and also that there are no states charged under both \U1s. If we were to diagonalise the \U1s throughout the RGE trajectory, we would introduce millicharges, and so to return to this basis at low energies, we would necessarily have to undo the transformation, affecting all of the parameters. Fortunately, maintaining in this basis with these assumptions greatly simplifies the RGEs, and it is easy to see that $\chi$ only appears as $\chi^2$. The largest effect is then induced through the visible sector terms; the leading corrections to the above RGEs are given by \cite{Morrissey:2009ur}
\begin{align}
\delta \left(\frac{d m_\pm^2}{dt}\right) =& - \frac{8 \chi^2 \alpha_h}{4\pi} |M_1|^2\nn\\
\delta \left(\frac{d A_S}{dt}\right) =& - \frac{8 \chi^2 \alpha_h}{4\pi} M_1
\end{align}
where $M_1$ is the Bino mass. When we recall equation (\ref{ChiRelation}), we see that this correction becomes
\begin{equation}
- \frac{8 \chi^2 \alpha_h}{4\pi} |M_1|^2 \rightarrow - \frac{8 \kappa^2 \alpha_h^2 \alpha_Y}{(4\pi)^3} |M_1|^2,
\end{equation}
i.e. it is a three-loop effect that can be safely neglected in the models we consider in this paper -- particularly since we assume gravity mediated generation of soft masses; these effects might be important (i.e. providing a leading but still very small contribution) if we were to take some of the them to be zero, for example, in little gauge mediation when the singlet and gaugino masses vanish.

%Finally we note that for the kinetic mixing, to all orders its RGE is
%\begin{align}
%\frac{d \chi}{dt} =& \frac{1}{2} \chi \bigg[ \alpha_h^{-1} \frac{d \alpha_h}{dt} + \alpha_Y^{-1} \frac{d \alpha_Y}{dt} \bigg]
%\end{align}

% ; for example, the hidden gauge coupling RGE is given to all orders by the expression
%\begin{align}
%\frac{d \alpha_h^{-1}}{dt} =& - \frac{2 - \gamma_{H_+} - \gamma_{H_-}}{2\pi}
%\end{align}
%where $\gamma_{H_+},\gamma_{H_-}$ are the anomalous dimensions of the fields $H_{\pm}$. The kinetic mixing appears in the anomalous dimensions of the fields as
%\begin{align}
%\gamma_{H_+} =&
%\end{align}

\section{Spectrum of the model}
\label{APP:LowE}

In this appendix, we present the details of the low-energy
features of the model $W = \lambda_S S H_+ H_-$. Once
supersymmetry and R-symmetry is broken, the potential is
\begin{align}
V =& |\lambda_S|^2 ( |S H_+|^2 + |S H_-|^2 + |H_+ H_-|^2) \nn\\
&+\frac{g^2_h}{2} (|H_+|^2 - |H_-|^2 - \xi)^2 \nn\\
&+m_+^2 |H_+|^2 + m_-^2 |H_-|^2 + m_S^2 |S|^2 \nn\\
&+ (\lambda_S A_S SH_+ H_- + c.c.)
\end{align}
We are assuming that no $\mu, B_\mu$ terms are generated; these
would introduce new scales into the theory. Although they could
conceivably be generated by a Giudice-Masiero mechanism in
analogy with the visible sector, we shall neglect this
possibility.

\subsection{Scalars}

Defining $\Delta \equiv \sqrt{\lambda_S^2 \xi - m_+^2
\lambda_S^2/g_h^2}$,
 we have mass matrices in the $(H_+,H_+^\dagger)$ basis of
\begin{align}
\frac{1}{2} (H_+^\dagger \; H_+) \2b2[ g^2 \Delta^2/\lambda_S^2, g^2_h \Delta^2/\lambda_S^2][g_h^2 \Delta^2/\lambda_S^2, g_h^2 \Delta^2/\lambda_S^2] \vec2[H_+,H_+^\dagger]
\end{align}
which implies masses for the two components of $\sqrt{2}g_h
\Delta/\lambda_S, 0$ at this level. The ``massless'' mode is
the Goldstone boson that becomes the longitudinal component of
the massive gauge field. The $(H_-,H_-^\dagger, S, S^\dagger) $
system is more complicated; we find a mass matrix of \beq
\frac{1}{2} (H_-^\dagger \; H_-\; S^\dagger\; S ) \left(
\begin{array}{cccc}
\Delta^2 + m_+^2 + m_-^2 &0&0&A_S^\dagger \Delta\\
0 &  \Delta^2 + m_+^2 + m_-^2 &A_S \Delta&0 \\
0 & A_S^\dagger \Delta& \Delta^2 + m_S^2 & 0 \\
A_S \Delta & 0& 0& \Delta^2 + m_S^2
\end{array} \right) \left(\begin{array}{c} H_-\\H_-^\dagger \\ S\\S^\dagger \end{array}\right).
\eeq

The above theory for non-zero $\xi$ has a minimum at $\bra H_+
\ket = \Delta/\lambda_S$ provided that $\Delta$ is real, and
\begin{align}
0 \le& m_-^2 + m_+^2 + m_S^2 + 2\Delta^2 \\
0 \le& ( m_-^2 + m_+^2 + \Delta^2) ( m_S^2 + \Delta^2 ) - |A_S|^2 \Delta^2. \nn
\end{align}
In the case $A_S =0, m_+^2 = m_-^2 < 0$, this translates simply
to the condition that $\lambda_S^2 \ge 2g_h^2$.

\subsection{Fermions and neutralino mixing}
\label{APP:FERMIONS}

The fermion mass matrix in the basis $(\tilde{\lambda},
\tilde{h}_+, \tilde{h}_-, \tilde{s} )$ (neglecting the kinetic
mixing of the gaugino with the neutralino) is given by
\begin{equation} \C{M}_f = \left(
\begin{array}{cccc} M_\lambda & m_{\gamma^\prime} & 0 & 0 \\
m_{\gamma^\prime} & 0 & 0 & 0 \\ 0 & 0 & 0 & \Delta \\
0&0&\Delta & 0 \end{array} \right).
\end{equation}
However, to properly compute the dark matter density,  we should
take mixing of the fermions with the neutralino into account.
The fields $\tilde{h}_-, \tilde{s}$ form a Dirac fermion that
does not mix with any other fields. There will, however, be
kinetic mixing of the Bino with the hidden gaugino, and
possibly mass mixing; writing these fields before the mixing as,
respectively, $\tilde{b}, \tilde{\lambda}$ we can define
\begin{align}
\tilde{b} =& \frac{1}{c_\epsilon}b \nn\\
\tilde{\lambda} =&  \lambda - t_\epsilon b,
\end{align}
which then allows us to write the full neutralino mass matrix
in the basis $(B_0, W_0, h_u^0, h_d^0, \tilde{\lambda},
\tilde{h}_+)$, including the standard Majorana masses for Bino
and Wino $M_{1,2}$ but also a potential explicit mass mixing
term $\C{L} \supset - m_X \tilde{b} \tilde{\lambda}$ as
 \begin{equation}
{
\left(\!\!\begin{array}{c c c c c c}
\frac{1}{c_\epsilon^2} [ M_1 - s_\epsilon m_X + s_\epsilon^2 M_\lambda ] & 0       & -M_Z s_W c_\beta/c_\epsilon &   M_Z s_W s_\beta/c_\epsilon &\frac{1}{c_\epsilon} [ m_X - s_\epsilon M_\lambda ]&-m_\gamma^\prime t_\epsilon \\
 0 & M_2   &  M_Z c_W c_\beta & - M_Z c_W s_\beta &0&0 \\
 -M_Z s_W c_\beta/c_\epsilon &   M_Z c_W c_\beta & 0    & -\tilde{\mu} &0&0\\
  M_Z s_W s_\beta/c_\epsilon &  -M_Z c_W s_\beta & -\tilde{\mu} & 0  &0&0  \\
\frac{1}{c_\epsilon} [ m_X - s_\epsilon M_\lambda ]&0&0&0&M_\lambda &m_\gamma^\prime\\
-m_\gamma^\prime t_\epsilon &0&0&0&m_\gamma^\prime&0
\end{array}\!\!\right) }
\end{equation}

\section{Kinetic and mass mixing}
\label{APP:KM}

Here, we review the diagonalisation of the gauge fields and the
subsequent coupling of the physical gauge bosons to matter
fields.

Consider the Lagrangian coupling the currents $j_B, j_W$ and $j_X$ to the respective unrotated gauge bosons $\tilde{B}_\mu, \tilde{W}_\mu, \tilde{X}_\mu$ (corresponding to hypercharge, weak and hidden gauge bosons):
\begin{align}
\C{L} =& - \frac{1}{4} \tilde{B}_{\mu\nu} \tilde{B}^{\mu\nu} - \frac{1}{4} \tilde{X}_{\mu\nu} \tilde{X}^{\mu\nu} + \frac{\chi}{2} \tilde{B}_{\mu\nu} \tilde{X}^{\mu\nu} - \frac{1}{4} \tilde{W}_{\mu\nu} \tilde{W}^{\mu\nu} + \frac{1}{2} \tilde{m}^2 \tilde{X}_{\mu} \tilde{X}^{\mu} + \frac{1}{8} v^2 (g_Y \tilde{B}_\mu - g_2 \tilde{W}_\mu)^2\nn\\
& + g_Y j^\mu_B \tilde{B}_\mu + g_2 j^\mu_W \tilde{W}_\mu+ g_h j_X^\mu \tilde{X}_\mu \nn\\
=& - \frac{1}{4} B_{\mu\nu} B^{\mu\nu} - \frac{1}{4} X_{\mu\nu} X^{\mu\nu}  - \frac{1}{4} \tilde{W}_{\mu\nu} \tilde{W}^{\mu\nu}  + g_2 j^\mu_W \tilde{W}_\mu + g_Y j^\mu_B B_\mu + \frac{1 }{\sqrt{1-\chi^2}} (g_h j_X^\mu +\chi g_Y j^\mu_B )X_\mu \nn\\
&+ \frac{\tilde{m}^2}{1-\chi^2}\frac{1}{2} X_{\mu} X^{\mu} + \frac{1}{8} v^2 (g_Y B_\mu + \frac{g_Y \chi}{\sqrt{1-\chi^2}} X_\mu- g_2 \tilde{W}_\mu)^2.
\end{align}
Then we make the transformation
\begin{align}
\tilde{W}_\mu \equiv& s_W A_\mu + c_W ( c_\phi Z_\mu + s_\phi \gamma^\prime_\mu)  \nn\\
\tilde{B}_\mu \equiv& c_W A_\mu - s_W (c_\phi Z_\mu + s_\phi \gamma^\prime_\mu) + \frac{\chi}{\sqrt{1-\chi^2}} (c_\phi \gamma^\prime_\mu - s_\phi Z_\mu) \nn\\
=& c_W A_\mu -( s_W c_\phi + \frac{\chi}{\sqrt{1-\chi^2}} s_\phi) Z_\mu + (\frac{c_\phi \chi}{\sqrt{1-\chi^2}} - s_W s_\phi)\gamma^\prime_\mu \nn\\
\tilde{X}_\mu\equiv&  \frac{1}{\sqrt{1-\chi^2}}(  - s_\phi
Z_\mu +c_\phi \gamma^\prime_\mu)
\label{ReparamGauge}\end{align} where $c_W, s_W$ are the usual
cosine and sine of the weak mixing angle, respectively, and $c_\phi,
s_\phi$ are the cosine and sine of an angle $\phi$ to be
determined below so that
\begin{align}
\C{L} \supset& - \frac{1}{4} F_{\mu\nu} F^{\mu\nu}  + m_{\gamma'}^2 \frac{1}{2} \gamma^\prime_{\mu} (\gamma^\prime)^{\mu} + M_Z^2 \frac{1}{2} Z_\mu Z^\mu\nn\\
& + e A_\mu \bigg[ j^\mu_W + j^\mu_B \bigg]  \nn\\
&+ Z_\mu\bigg[  g_2 c_W  c_\phi j^\mu_W -(  s_W c_\phi + \frac{ \chi s_\phi }{\sqrt{1-\chi^2}} )g_Yj^\mu_B -\frac{g_X s_\phi}{\sqrt{1-\chi^2}} j_X^\mu \bigg]\nn\\
&+ \gamma^\prime_\mu \bigg[ g_2 c_W s_\phi  j^\mu_W  + (\frac{c_\phi \chi}{\sqrt{1-\chi^2}} -  s_W s_\phi) g_Y j^\mu_B +\frac{g_X c_\phi}{\sqrt{1-\chi^2}} j_X^\mu \bigg].
\label{EQ:PhysicalCouplings}\end{align}
We find, defining $x\equiv \tilde{m}^2/M_Z^2 (\simeq m_{\gamma'}^2/M_Z^2)$,
\begin{align}
\tan 2 \phi =& - \frac{ 2 s_W s_\epsilon c_\epsilon }{c_\epsilon^2 - s_W^2 s_\epsilon^2 - x} \nn\\
\sin \phi =& \frac{s_W \chi}{1-x} + ...
\end{align}
where $s_\epsilon = -\chi, c_\epsilon = \sqrt{1-\chi^2}$ were defined in the text.

In terms of these, we have the eigenvalues where $m_+$
corresponds to the physical $Z$ mass and $m_-$ to the physical
hidden photon mass:
\begin{align}
m_\pm^2 =&\frac{1}{2} \bigg[ m^2 + \frac{M_Z^2}{\cos^2 \alpha} \pm \sqrt{(\frac{M_Z^2}{\cos^2\alpha} + m^2)^2 - 4 m^2 M_Z^2}\bigg] \nn\\
m_+^2 =& M_Z^2 \bigg[ 1 + \frac{s_W^2 \chi^2}{1-x} + ... \bigg] \nn\\
m_-^2 =& \tilde{m}^2 \bigg[ 1 + \frac{(1- s_W^2 - x)\chi^2}{1-x} + ... \bigg].
\end{align}
Thus the masses are only shifted at order $\chi^2$.

Note that we can also write
\begin{align}
m_+^2 =& M_Z^2 (c_\phi - s_W t_\epsilon s_\phi)^2 + \frac{\tilde{m}^2}{c_\epsilon^2} s_\phi^2 \nn\\
m_-^2 =& M_Z^2 (s_\phi + s_W t_\epsilon c_\phi)^2 + \frac{\tilde{m}^2}{c_\epsilon^2} c_\phi^2
\end{align}
and, defining $\hat{x} \equiv m_-^2/m_+^2 \approx x$, we have
\begin{align}
\tan \phi =& \frac{- (1 - \hat{x}) \pm \sqrt{(1 - \hat{x})^2 - 4 s_W^2 t_\epsilon^2 x}}{2s_W t_\epsilon \hat{x}}.
\label{tanphi}\end{align}

\section{Goldstone boson mixing}
\label{APP:GOLDSTONE}

Here, we consider the effect of the mixing on the Goldstone
bosons that are eaten and what happens to the other fields.
Assuming that the visible sector is the MSSM with neutral Goldstone boson $G^0$, and taking the
hidden sector to be broken by a single complex scalar $C = \frac{1}{\sqrt{2}} (C_R + i C_I)$, we
have
\begin{align}
\L ~\supset~& | \partial_\mu C - i g_h \tilde{X}_\mu C|^2 + |\partial_\mu H_u^0 -\frac{i}{2} (g_Y B_\mu - g_2 W_\mu^0)H_u^0|^2 + |\partial_\mu H_d^0 + \frac{i}{2} (g_Y B_\mu - g_2 W_\mu^0)H_d^0|^2 \nn\\
&+ V_{hid} (C) + V_{vis} (H)\nn\\
~=~& \frac{1}{2}(\partial_\mu C_I - m_h \tilde{X}_\mu )^2 + \frac{1}{2} (\partial_\mu C_R)^2 + \frac{1}{2}\bigg[\partial_\mu (c_\beta H_3 - s_\beta G^0) -\frac{1}{2} (g_Y B_\mu - g_2 W_\mu^0)vs_\beta \bigg]^2\nn\\
&+\frac{1}{2}\bigg[\partial_\mu (s_\beta H_3 + c_\beta G^0) + \frac{1}{2} (g_Y B_\mu - g_2 W_\mu^0) vc_\beta \bigg]^2 + ... \nn\\
~\supset~& \frac{1}{2}(\partial_\mu C_I - m_h \tilde{X}_\mu )^2 + \frac{1}{2} \bigg[\partial_\mu G^0 - \frac{ev}{s_{2W}} (c_W W_\mu^0 - s_W B_\mu) \bigg]^2 + ...
\end{align}
Clearly, the masses of the neutral Higgs (both visible and
hidden) are unaffected by the mixing, but the pseudoscalar will
be. Using (\ref{ReparamGauge}) and
\begin{align}
 c_W W_\mu^0 - s_W B_\mu =& (c_\phi - s_W t_\epsilon s_\phi) Z_\mu + (s_\phi + s_W t_\epsilon c_\phi) \gamma_\mu^\prime
\end{align}
we obtain
\begin{align}
\C{L} ~\supset~& - m_h \partial_\mu C_I \frac{1}{c_\epsilon}(  - s_\phi Z_\mu +c_\phi \gamma^\prime_\mu) - M_{Z^0} \partial_\mu G^0 ((c_\phi - s_W t_\epsilon s_\phi) Z_\mu + (s_\phi + s_W t_\epsilon c_\phi) \gamma_\mu^\prime)\nn\\
~\supset~& - Z_\mu \bigg( -m_h  \frac{s_\phi}{c_\epsilon} \partial_\mu C_I + M_{Z^0} (c_\phi - s_W t_\epsilon s_\phi) \partial_\mu G^0\bigg) \nn \\
& - \gamma^\prime_\mu \bigg( m_h \frac{c_\phi}{c_\epsilon}\partial_\mu C_I +  M_{Z^0} (s_\phi + s_W t_\epsilon c_\phi)\partial_\mu G^0 \bigg)
\end{align}
Thus,
\begin{align}
m_+  G_Z =& M_{Z^0} (c_\phi - s_W t_\epsilon s_\phi) G^0 -m_h  \frac{s_\phi}{c_\epsilon}  C_I \nn\\
m_-  G_{\gamma'} =& M_{Z^0} (s_\phi + s_W t_\epsilon c_\phi) G^0 + m_h \frac{c_\phi}{c_\epsilon} C_I
\end{align}
and
\begin{align}
G^0 =& \frac{c_\epsilon}{M_{Z^0} m_h}\bigg[ \frac{m_h c_\phi m_+}{c_\epsilon} G_Z + \frac{m_h s_\phi m_-}{c_\epsilon}G_{\gamma'} \bigg] \nn\\
=& \frac{1}{M_{Z^0}} \bigg[ c_\phi m_+ G_Z +  s_\phi m_- G_{\gamma'} \bigg] \nn\\
C_I =& \frac{c_\epsilon}{m_h} \bigg[ -(s_\phi + s_W t_\epsilon c_\phi)  m_+ G_Z + (c_\phi- s_W t_\epsilon s_\phi) m_- G_{\gamma'} \bigg].
\end{align}
We can also write this as
\begin{align}
G^0 =& \cos \psi G_Z + \sin \psi G_{\gamma'}  \nn\\
C_I =& -\sin \psi G_Z + \cos \psi G_{\gamma'} \nn\\
\tan \psi =& (\tan \phi) \frac{m_-}{m_+} = \frac{ (s_\phi + s_W t_\epsilon c_\phi)  m_+}{(c_\phi- s_W t_\epsilon s_\phi) m_-}.
\end{align}

\section{Higgs portal mixing}
\label{APP:HIGGS}

With the Lagrangian density
\begin{equation}
 \C{L} \supset \int d^2 \theta
\bigg( \frac{1}{4} B^\alpha B_\alpha + \frac{1}{4}
X^\alpha X_\alpha - \frac{\chi}{2} B^\alpha X_\alpha \bigg) +
c.c.,
\end{equation}
we need the $D$-term mixing. We write $\hat{D}_Y \equiv - g_Y \sum_{\phi} \phi^\dagger \hat{Y} \phi, \hat{D}_X \equiv -g_h \sum_\phi \phi^\dagger \hat{Q}_X \phi$ as the D-terms in the absence of mixing; then, we have
\begin{align}
\C{L} \supset& \frac{1}{2} D_Y^2 + \frac{1}{2} D_X^2 -  \chi D_Y D_X - D_Y \hat{D}_Y -  D_X \hat{D}_X
\end{align}
which leads to
\begin{align}
D_X =&  \frac{1}{1-\chi^2} ( \hat{D}_X + \chi \hat{D}_Y) \nn\\
D_Y =& \frac{1}{1-\chi^2} ( \hat{D}_Y + \chi \hat{D}_X)
\end{align}
and thus
\begin{align}
V \rightarrow& \frac{1}{2}\frac{1}{1-\chi^2} \bigg[ \hat{D}_X^2 + \hat{D}_Y^2 + 2\chi \hat{D}_X \hat{D}_Y \bigg], \nn\\
V_{\mathrm{Portal}} \equiv& \frac{\chi}{1-\chi^2}\hat{D}_X \hat{D}_Y.
\end{align}
The relevant part of the potential for us involves the Higgses;
we can write the portal term as
\begin{alignat}{3}
V_{\mathrm{Portal}} ~=~& \frac{\chi}{1-\chi^2} g_Y g_h (|H_+|^2 - |H_-|^2) &&( \frac{1}{2} |H_u|^2 - \frac{1}{2} |H_d|^2 )\\
~=~&   \frac{\chi}{1-\chi^2} g_Y g_h (|H_+|^2 - |H_-|^2) &&\bigg[ \frac{1}{2} |H_u^+|^2 +  \frac{1}{2} \bigg( \frac{1}{\sqrt{2}} v\sin \beta + H_u^0 \bigg)^2 \nn \\
& && -  \frac{1}{2} |H_d^-|^2 -  \frac{1}{2} \bigg( \frac{1}{\sqrt{2}} v\cos \beta + H_d^0 \bigg)^2 \bigg]. \nn
\end{alignat}
Immediately, we can extract the effective Fayet-Iliopoulos term:
\begin{align}
V_{FI} =& \frac{g_h^2}{2} (|H_+|^2 - |H_-|^2 - \xi)^2 \nn\\
=& \frac{1}{2} (\hat{D}_h + \xi/g_h)^2 \nn\\
\xi \approx& (\chi/ g_h) \bra \hat{D}_Y\ket \nn\\
=& (\chi/g_h) g_Y \frac{v^2}{4} \cos 2 \beta.
\end{align}
However, we can also extract the Higgs mass mixing. Writing
\begin{align}
H_+ =& \frac{v_+}{\sqrt{2}} + \frac{1}{\sqrt{2}} (x_R + i x_I) \nn\\
H_u^0 =& \frac{1}{\sqrt{2}} [ s_\beta v + h_u^0  + i (c_\beta A - s_\beta G^0 )]\nn\\
H_d^0 =& \frac{1}{\sqrt{2}} [ c_\beta v + h_d^0  + i (s_\beta A + c_\beta G^0)] \nn\\
\vec2[ h_d^0,h_u^0] =& \arr2[\cos \alpha, -\sin\alpha][\sin \alpha, \cos \alpha] \vec2[H, h]
\end{align}
and using the standard shorthand $c_\beta \equiv \cos \beta,
c_\alpha \equiv \cos \alpha, c_{\alpha + \beta} \equiv \cos
(\alpha + \beta)$, etc., we have
\begin{align}
V \supset& -\frac{t_\epsilon}{c_\epsilon} g_Y g_h |\frac{v_+}{\sqrt{2}} + \frac{1}{\sqrt{2}} (x_R + i x_I)|^2 \nn\\
&\times\bigg[\frac{1}{2} |\frac{1}{\sqrt{2}} [ s_\beta v + h_u^0  + i (c_\beta A - s_\beta G^0 )]|^2 - \frac{1}{2}  |\frac{1}{\sqrt{2}} [ c_\beta v + h_d^0 + i (s_\beta A + c_\beta G^0)]|^2 \bigg]\nn\\
\supset& -\frac{t_\epsilon}{c_\epsilon} g_Y g_h \frac{1}{2} v_+ v x_R \bigg[ s_{\alpha + \beta} h - c_{\alpha + \beta} H \bigg] \nn\\
\equiv& \frac{1}{2} M_m^2 x_R \bigg[ - s_{\alpha + \beta} h + c_{\alpha + \beta} H \bigg] \nn\\
M_m^2 \equiv& \frac{t_\epsilon}{c_\epsilon} g_Y g_h  v \frac{\sqrt{2} \Delta}{\lambda_S} \nn\\
=&\frac{t_\epsilon}{c_\epsilon} 2M_Z s_W m_{\gamma^\prime} \approx - 2\chi s_W M_Z m_{\gamma^\prime}.
\end{align}
So then we must redefine our Higgses: the mass mixing matrix in
the basis $(x_+, h, H)$ is
\begin{equation}
\mathcal{M}_{\mathrm{Higgs}}^2 = \left( \begin{array}{ccc} m_{+}^2 & - s_{\alpha + \beta} M_m^2 & c_{\alpha + \beta} M_m^2 \\
 - s_{\alpha + \beta} M_m^2 & m_h^2 & 0 \\
c_{\alpha + \beta} M_m^2 & 0 & m_H^2 \end{array} \right).
\end{equation}
To first order in $M_m^2/m_{h,H}^2$, this is diagonalised via
\begin{equation}
\left( \begin{array}{c} x_+ \\ h \\ H \end{array} \right) = \left( \begin{array}{ccc} 1 & - \frac{s_{\alpha + \beta} M_m^2}{m_h^2 - m_+^2} & \frac{c_{\alpha + \beta} M_m^2}{m_H^2 - m_+^2} \\
 \frac{s_{\alpha + \beta} M_m^2}{m_h^2 - m_+^2} &1 & 0 \\
-\frac{c_{\alpha + \beta} M_m^2}{m_H^2 - m_+^2} & 0 & 1 \end{array} \right) \left( \begin{array}{c} x_R^\prime \\ h^\prime \\ H^\prime \end{array} \right).
\end{equation}

Since we often find $m_H^2 \gg m_h^2$, however, the above will
usually reduce to mixing between the hidden and lightest Higgs.
In this case, we can approximate
\begin{align}
\vec2[x_R, h] \approx& \2b2[1, u][-u,1] \vec2[x_R^\prime,h^\prime] \nn\\
u \approx& \frac{-s_{\alpha + \beta}M_m^2 }{m_h^2 - m_+^2} \approx s_{\alpha + \beta} 2\chi s_W \frac{M_Z m_{\gamma^\prime}}{m_h^2}.
\end{align}
Thus, very roughly, $u \sim \chi s_W m_{\gamma^\prime}/m_h$ for
large $\tan \beta$.

\subsection{Spin-independent nucleon cross sections}

Here, we would like to estimate the cross sections for our
Majorana hidden fermion $\tilde{O}_1 = \vec2[\tilde{o}_1,
\ov{\tilde{o}}_1]$ on nucleons that take place via the Higgs
portal term. The Higgs portal leads to an effective four-point
interaction
\begin{align}
\C{L} \supset f_N \bigg(\ov{\tilde{O}}_1 \tilde{O}_1 \ov{N} N\bigg)
\end{align}
and we can consider different $f_p, f_n$ for protons and
neutrons, respectively. Consider that the dark matter particle
is a Majorana combination of $\tilde{\lambda}$ and
$\tilde{h}_+$ fermions, $\tilde{o}_1 \approx \cos \theta_1
\tilde{h}_+ + \sin \theta_1 \tilde{\lambda}$. The coupling to
the hidden Higgs is via the kinetic vertex
\begin{align}
\C{L} \supset& - \sqrt{2} g_h H_+^* (\tilde{h}_+ \tilde{\lambda}) + c.c.\nn\\
\supset& - g_h  \cos \theta_1 \sin \theta_1 x_R [(\tilde{o}_1 \tilde{o}_1 ) + (\ov{\tilde{o}}_1 \ov{\tilde{o}}_1) ] \nn\\
\supset&  - g_h  \cos \theta_1 \sin \theta_1 x_R \ov{\tilde{O}}_1 \tilde{O}_1
\end{align}
where now $\tilde{O}_1$ is in Dirac form, $\tilde{O}_1 =
\vec2[\tilde{o}_1, \ov{\tilde{o}}_1]$. Now let us write the
coupling of the MSSM Higgs to nucleons as $- a_N h \ov{N} N$.
\begin{align}
\C{L} \supset&  - g_h  \cos \theta_1 \sin \theta_1 [ x_R^\prime + u h^\prime] \ov{\tilde{O}}_1 \tilde{O}_1 - a_N [ -u x_R^\prime + h^\prime] \ov{N} N - \frac{1}{2} m_1^2 (x^\prime_R)^2 - \frac{1}{2} m_2^2 (h^\prime)^2  \nn\\
\rightarrow& u a_N \cos\theta_1 \sin\theta_1  \bigg[ \frac{1}{m_1^2} - \frac{1}{m_2^2}\bigg] \bigg(\ov{\tilde{O}}_1 \tilde{O}_1 \ov{N} N\bigg) \nn\\
\approx& a_N \cos\theta_1 \sin\theta_1 s_{\alpha + \beta} 2\chi s_W \frac{M_Z m_{\gamma^\prime}}{m_+^2 m_h^2} \bigg(\ov{\tilde{O}}_1 \tilde{O}_1 \ov{N} N\bigg) \nn\\
\rightarrow f_N \approx& a_N \sin 2\theta_1 s_{\alpha + \beta} \chi s_W \frac{M_Z m_{\gamma^\prime}}{m_+^2 m_h^2}.
\end{align}
So then, our direct detection amplitude should be approximately
given by the coefficient above.

The coupling of the MSSM Higgs to nucleons is determined by its
coupling to quarks. These come from
\begin{align}
\C{L} \supset& - Y_U H_U qu - Y_D H_D qd \nn\\
\supset& - \frac{c_\alpha}{ v s_\beta} h ( m_U \ov{u} u) - \frac{s_\alpha}{ v c_\beta} h (m_D \ov{d} d).
\end{align}
Then, we use that \cite{Jungman:1995df}
\begin{align}
\bra N | m_q \ov{q} q | N \ket =& m_n f_{Tq}^{(N)} \nn\\
\bra N | m_Q \ov{Q} Q | N \ket_{Q=c,b,t} =& \frac{2}{27} m_N \bigg[ 1 - \sum_{q=u,d,s} f_{Tq}^{(N)} \bigg] \nn\\
\equiv& \frac{2}{27} m_N \bigg[ 1 - \tilde{F}^{(N)} \bigg]
\end{align}
to give
\begin{align}
a_N =& \frac{c_\alpha}{ v s_\beta} m_N \bigg[ \frac{4}{27} ( 1 - \tilde{F}) + f_{Tu}^{(N)} \bigg] + \frac{s_\alpha}{ v c_\beta} m_N \bigg[ \frac{2}{27} ( 1 - \tilde{F}) +   f_{Td}^{(N)} + f_{Ts}^{(N)}\bigg].
\end{align}

If we consider large $\tan \beta$ with $\beta = \pi/2 -
\delta$, $\alpha \approx - \delta$, we have $s_\alpha \approx -
\delta, c_\alpha \approx 1, c_\beta \approx \delta, s_\beta
\approx 1$ and so $a_N \approx \frac{m_N}{v} \bigg[
\frac{2}{27} ( 1 - \tilde{F}) + f_{Tu}^{(N)} -  f_{Td}^{(N)} -
f_{Ts}^{(N)}\bigg]$. Let us then simply define
\begin{align}
a_N \equiv \frac{m_N}{v} \hat{f}^{(N)}.
\end{align}
There are large uncertainties in the value of $\hat{f}^{(N)}$,
however, it is approximately equal for protons and neutrons and
varies from about $0.03$ to $0.44$. We can then take an
approximate value for the amplitude for a Majorana fermion
scattering on nucleons to be
\begin{align}
f_N \approx \sin 2\theta_1 s_{\alpha + \beta} \chi s_W \frac{M_Z m_{\gamma^\prime}}{m_+^2 m_h^2} \frac{m_N}{v} \hat{f}^{(N)}.
\end{align}
Taking the large $\tan \beta $ values, $m_h = 115$ GeV and
$\hat{f}^{(N)} \sim 0.1$ we obtain
\begin{align}
f_N \approx 3\times 10^{-9} (\mathrm{GeV})^{-2}\times \bigg( \sin 2\theta_1 \bigg) \left( \frac{\chi s_W}{0.001} \right)  \left(\frac{\mathrm{GeV}}{m_{\gamma'}} \right)
\end{align}
This is clearly a very small effect. This translates into a
cross section for scattering on a single nucleon of
\begin{align}
\sigma_N^{\rm SI,\,Portal} =& \frac{4 m_{\tilde{o}1}^2 m_N^2}{\pi (m_N +  m_{\tilde{o}1})^2} f_N^2 \nn\\
\approx& 2 \times 10^{-45} \mathrm{cm}^2 \times \left(\frac{m_{\tilde{o}1}^2 }{(m_N +  m_{\tilde{o}1})^2}\right)\bigg( \sin 2\theta_1 \bigg)^2 \left( \frac{\chi s_W}{0.001} \right)^2  \left(\frac{\mathrm{GeV}}{m_{\gamma'}} \right)^2.
\end{align}
This corresponds well to the values that we find in the plots (see figure \ref{VisibleSI}). %\marginnote{fig 11 $~~$} %

The above will be supplemented by contributions from
$s$-channel squark exchange. Very roughly for these, we have
effective four-point interactions of the Majorana fermion with
quarks with coupling $ f_q \sim \frac{g^2}{M_Q^2} |U_{\tilde{b}
\tilde{o}_1}|^2, $ where $\tilde{b} = \sum_n U_{b \tilde{o}_n}
\tilde{o}_n$ is the Bino, which mixes most strongly with the
lightest hidden state. By considering the mass mixing matrix in
section \ref{APP:FERMIONS}, we can conclude that (in the absence
of direct mass mixing) the mixing is simply $\sim \chi
M_\lambda$, so we have $U_{\tilde{b} \tilde{o}_1} \sim \chi$,
and thus
\begin{align}
f_N \sim& \frac{g^2_Y}{M_Q^2} \chi^2 \sum_q \frac{m_N}{m_q} f_{Tq}^{N}
\end{align}
and thus
\begin{align}
\sigma_N^{\rm SD,\,Squark} \sim& 10^{-48} \mathrm{cm}^2 \times \left(\frac{m_{\tilde{o}_1}^2 }{(m_N +  m_{\tilde{o}_1})^2}\right) \left( \frac{s_W\chi}{0.001} \right)^4  \left(\frac{100 \mathrm{GeV}}{M_{\tilde{Q}}} \right)^4 .
\end{align}

\small
\bibliographystyle{JHEP-2}
\addcontentsline{toc}{section}{References}
\bibliography{AGR}

\end{document}